\documentclass{article}
\usepackage{eepic,epic}
\usepackage{amsmath}
\usepackage{amssymb}

\def\PsfigVersion{1.9}
\ifx\undefined\psfig\else\endinput\fi

\let\LaTeXAtSign=\@
\let\@=\relax
\edef\psfigRestoreAt{\catcode`\@=\number\catcode`@\relax}
\catcode`\@=11\relax
\newwrite\@unused
\def\ps@typeout#1{{\let\protect\string\immediate\write\@unused{#1}}}
\ps@typeout{psfig/tex \PsfigVersion}

\def\figurepath{./}

\def\@nnil{\@nil}
\def\@empty{}
\def\@psdonoop#1\@@#2#3{}
\def\@psdo#1:=#2\do#3{\edef\@psdotmp{#2}\ifx\@psdotmp\@empty \else
    \expandafter\@psdoloop#2,\@nil,\@nil\@@#1{#3}\fi}
\def\@psdoloop#1,#2,#3\@@#4#5{\def#4{#1}\ifx #4\@nnil \else
       #5\def#4{#2}\ifx #4\@nnil \else#5\@ipsdoloop #3\@@#4{#5}\fi\fi}
\def\@ipsdoloop#1,#2\@@#3#4{\def#3{#1}\ifx #3\@nnil 
       \let\@nextwhile=\@psdonoop \else
      #4\relax\let\@nextwhile=\@ipsdoloop\fi\@nextwhile#2\@@#3{#4}}
\def\@tpsdo#1:=#2\do#3{\xdef\@psdotmp{#2}\ifx\@psdotmp\@empty \else
    \@tpsdoloop#2\@nil\@nil\@@#1{#3}\fi}
\def\@tpsdoloop#1#2\@@#3#4{\def#3{#1}\ifx #3\@nnil 
       \let\@nextwhile=\@psdonoop \else
      #4\relax\let\@nextwhile=\@tpsdoloop\fi\@nextwhile#2\@@#3{#4}}
\ifx\undefined\fbox
\newdimen\fboxrule
\newdimen\fboxsep
\newdimen\ps@tempdima
\newbox\ps@tempboxa
\long\def\fbox#1{\leavevmode\setbox\ps@tempboxa\hbox{#1}\ps@tempdima\fboxrule
    \advance\ps@tempdima \fboxsep \advance\ps@tempdima \dp\ps@tempboxa
   \hbox{\lower \ps@tempdima\hbox
  {\vbox{\hrule height \fboxrule
          \hbox{\vrule width \fboxrule \hskip\fboxsep
          \vbox{\vskip\fboxsep \box\ps@tempboxa\vskip\fboxsep}\hskip 
                 \fboxsep\vrule width \fboxrule}
                 \hrule height \fboxrule}}}}
\fi
\newread\ps@stream
\newif\ifnot@eof       %
\newif\if@noisy        %
\newif\if@atend        %
\newif\if@psfile       %
{\catcode`\%=12\global\gdef\epsf@start{
\def\epsf@PS{PS}
\def\epsf@getbb#1{%
\openin\ps@stream=#1
\ifeof\ps@stream\ps@typeout{Error, File #1 not found}\else
   {\not@eoftrue \chardef\other=12
    \def\do##1{\catcode`##1=\other}\dospecials \catcode`\ =10
    \loop
       \if@psfile
          \read\ps@stream to \epsf@fileline
       \else{
          \obeyspaces
          \read\ps@stream to \epsf@tmp\global\let\epsf@fileline\epsf@tmp}
       \fi
       \ifeof\ps@stream\not@eoffalse\else
       \if@psfile\else
       \expandafter\epsf@test\epsf@fileline:. \\%
       \fi
          \expandafter\epsf@aux\epsf@fileline:. \\%
       \fi
   \ifnot@eof\repeat
   }\closein\ps@stream\fi}%
\long\def\epsf@test#1#2#3:#4\\{\def\epsf@testit{#1#2}
                        \ifx\epsf@testit\epsf@start\else
\ps@typeout{Warning! File does not start with `\epsf@start'.  It may not be a PostScript file.}
                        \fi
                        \@psfiletrue} %
{\catcode`\%=12\global\let\epsf@percent=
\long\def\epsf@aux#1#2:#3\\{\ifx#1\epsf@percent
   \def\epsf@testit{#2}\ifx\epsf@testit\epsf@bblit
        \@atendfalse
        \epsf@atend #3 . \\%
        \if@atend       
           \if@verbose{
                \ps@typeout{psfig: found `(atend)'; continuing search}
           }\fi
        \else
        \epsf@grab #3 . . . \\%
        \not@eoffalse
        \global\no@bbfalse
        \fi
   \fi\fi}%
\def\epsf@grab #1 #2 #3 #4 #5\\{%
   \global\def\epsf@llx{#1}\ifx\epsf@llx\empty
      \epsf@grab #2 #3 #4 #5 .\\\else
   \global\def\epsf@lly{#2}%
   \global\def\epsf@urx{#3}\global\def\epsf@ury{#4}\fi}%
\def\epsf@atendlit{(atend)} 
\def\epsf@atend #1 #2 #3\\{%
   \def\epsf@tmp{#1}\ifx\epsf@tmp\empty
      \epsf@atend #2 #3 .\\\else
   \ifx\epsf@tmp\epsf@atendlit\@atendtrue\fi\fi}

\chardef\psletter = 11 %
\chardef\other = 12

\newif \ifdebug %
\newif\ifc@mpute %
\c@mputetrue %

\let\then = \relax
\def\r@dian{pt }
\let\r@dians = \r@dian
\let\dimensionless@nit = \r@dian
\let\dimensionless@nits = \dimensionless@nit
\def\internal@nit{sp }
\let\internal@nits = \internal@nit
\newif\ifstillc@nverging
\def \Mess@ge #1{\ifdebug \then \message {#1} \fi}

{ %
        \catcode `\@ = \psletter
        \gdef \nodimen {\expandafter \n@dimen \the \dimen}
        \gdef \term #1 #2 #3%
               {\edef \t@ {\the #1}%
                \edef \t@@ {\expandafter \n@dimen \the #2\r@dian}%
                \t@rm {\t@} {\t@@} {#3}%
               }
        \gdef \t@rm #1 #2 #3%
               {{%
                \count 0 = 0
                \dimen 0 = 1 \dimensionless@nit
                \dimen 2 = #2\relax
                \Mess@ge {Calculating term #1 of \nodimen 2}%
                \loop
                \ifnum  \count 0 < #1
                \then   \advance \count 0 by 1
                        \Mess@ge {Iteration \the \count 0 \space}%
                        \Multiply \dimen 0 by {\dimen 2}%
                        \Mess@ge {After multiplication, term = \nodimen 0}%
                        \Divide \dimen 0 by {\count 0}%
                        \Mess@ge {After division, term = \nodimen 0}%
                \repeat
                \Mess@ge {Final value for term #1 of 
                                \nodimen 2 \space is \nodimen 0}%
                \xdef \Term {#3 = \nodimen 0 \r@dians}%
                \aftergroup \Term
               }}
        \catcode `\p = \other
        \catcode `\t = \other
        \gdef \n@dimen #1pt{#1} %
}

\def \Divide #1by #2{\divide #1 by #2} %

\def \Multiply #1by #2%
       {{%
        \count 0 = #1\relax
        \count 2 = #2\relax
        \count 4 = 65536
        \Mess@ge {Before scaling, count 0 = \the \count 0 \space and
                        count 2 = \the \count 2}%
        \ifnum  \count 0 > 32767 %
        \then   \divide \count 0 by 4
                \divide \count 4 by 4
        \else   \ifnum  \count 0 < -32767
                \then   \divide \count 0 by 4
                        \divide \count 4 by 4
                \else
                \fi
        \fi
        \ifnum  \count 2 > 32767 %
        \then   \divide \count 2 by 4
                \divide \count 4 by 4
        \else   \ifnum  \count 2 < -32767
                \then   \divide \count 2 by 4
                        \divide \count 4 by 4
                \else
                \fi
        \fi
        \multiply \count 0 by \count 2
        \divide \count 0 by \count 4
        \xdef \product {#1 = \the \count 0 \internal@nits}%
        \aftergroup \product
       }}

\def\r@duce{\ifdim\dimen0 > 90\r@dian \then   %
                \multiply\dimen0 by -1
                \advance\dimen0 by 180\r@dian
                \r@duce
            \else \ifdim\dimen0 < -90\r@dian \then  %
                \advance\dimen0 by 360\r@dian
                \r@duce
                \fi
            \fi}

\def\Sine#1%
       {{%
        \dimen 0 = #1 \r@dian
        \r@duce
        \ifdim\dimen0 = -90\r@dian \then
           \dimen4 = -1\r@dian
           \c@mputefalse
        \fi
        \ifdim\dimen0 = 90\r@dian \then
           \dimen4 = 1\r@dian
           \c@mputefalse
        \fi
        \ifdim\dimen0 = 0\r@dian \then
           \dimen4 = 0\r@dian
           \c@mputefalse
        \fi
        \ifc@mpute \then
                \divide\dimen0 by 180
                \dimen0=3.141592654\dimen0
                \dimen 2 = 3.1415926535897963\r@dian %
                \divide\dimen 2 by 2 %
                \Mess@ge {Sin: calculating Sin of \nodimen 0}%
                \count 0 = 1 %
                \dimen 2 = 1 \r@dian %
                \dimen 4 = 0 \r@dian %
                \loop
                        \ifnum  \dimen 2 = 0 %
                        \then   \stillc@nvergingfalse 
                        \else   \stillc@nvergingtrue
                        \fi
                        \ifstillc@nverging %
                        \then   \term {\count 0} {\dimen 0} {\dimen 2}%
                                \advance \count 0 by 2
                                \count 2 = \count 0
                                \divide \count 2 by 2
                                \ifodd  \count 2 %
                                \then   \advance \dimen 4 by \dimen 2
                                \else   \advance \dimen 4 by -\dimen 2
                                \fi
                \repeat
        \fi             
                        \xdef \sine {\nodimen 4}%
       }}

\def\Cosine#1{\ifx\sine\UnDefined\edef\Savesine{\relax}\else
                             \edef\Savesine{\sine}\fi
        {\dimen0=#1\r@dian\advance\dimen0 by 90\r@dian
         \Sine{\nodimen 0}
         \xdef\cosine{\sine}
         \xdef\sine{\Savesine}}}              

\def\psdraft{
        \def\@psdraft{0}
}
\def\psfull{
        \def\@psdraft{100}
}

\psfull

\newif\if@scalefirst
\def\psscalefirst{\@scalefirsttrue}
\def\psrotatefirst{\@scalefirstfalse}
\psrotatefirst

\newif\if@draftbox
\def\psnodraftbox{
        \@draftboxfalse
}
\def\psdraftbox{
        \@draftboxtrue
}
\@draftboxtrue

\newif\if@prologfile
\newif\if@postlogfile
\def\pssilent{
        \@noisyfalse
}
\def\psnoisy{
        \@noisytrue
}
\psnoisy
\newif\if@bbllx
\newif\if@bblly
\newif\if@bburx
\newif\if@bbury
\newif\if@height
\newif\if@width
\newif\if@rheight
\newif\if@rwidth
\newif\if@angle
\newif\if@clip
\newif\if@verbose
\def\@p@@sclip#1{\@cliptrue}

\newif\if@decmpr

\def\@p@@sfigure#1{\def\@p@sfile{null}\def\@p@sbbfile{null}
                \openin1=#1.bb
                \ifeof1\closein1
                        \openin1=\figurepath#1.bb
                        \ifeof1\closein1
                                \openin1=#1
                                \ifeof1\closein1%
                                       \openin1=\figurepath#1
                                        \ifeof1
                                           \ps@typeout{Error, File #1 not found}
                                                \if@bbllx\if@bblly
                                                \if@bburx\if@bbury
                                                        \def\@p@sfile{#1}%
                                                        \def\@p@sbbfile{#1}%
                                                        \@decmprfalse
                                                \fi\fi\fi\fi
                                        \else\closein1
                                                \def\@p@sfile{\figurepath#1}%
                                                \def\@p@sbbfile{\figurepath#1}%
                                                \@decmprfalse
                                        \fi%
                                \else\closein1%
                                        \def\@p@sfile{#1}
                                        \def\@p@sbbfile{#1}
                                        \@decmprfalse
                                \fi
                        \else
                                \def\@p@sfile{\figurepath#1}
                                \def\@p@sbbfile{\figurepath#1.bb}
                                \@decmprtrue
                        \fi
                \else
                        \def\@p@sfile{#1}
                        \def\@p@sbbfile{#1.bb}
                        \@decmprtrue
                \fi}

\def\@p@@sfile#1{\@p@@sfigure{#1}}

\def\@p@@sbbllx#1{
                \@bbllxtrue
                \dimen100=#1
                \edef\@p@sbbllx{\number\dimen100}
}
\def\@p@@sbblly#1{
                \@bbllytrue
                \dimen100=#1
                \edef\@p@sbblly{\number\dimen100}
}
\def\@p@@sbburx#1{
                \@bburxtrue
                \dimen100=#1
                \edef\@p@sbburx{\number\dimen100}
}
\def\@p@@sbbury#1{
                \@bburytrue
                \dimen100=#1
                \edef\@p@sbbury{\number\dimen100}
}
\def\@p@@sheight#1{
                \@heighttrue
                \dimen100=#1
                \edef\@p@sheight{\number\dimen100}
}
\def\@p@@swidth#1{
                \@widthtrue
                \dimen100=#1
                \edef\@p@swidth{\number\dimen100}
}
\def\@p@@srheight#1{
                \@rheighttrue
                \dimen100=#1
                \edef\@p@srheight{\number\dimen100}
}
\def\@p@@srwidth#1{
                \@rwidthtrue
                \dimen100=#1
                \edef\@p@srwidth{\number\dimen100}
}
\def\@p@@sangle#1{
                \@angletrue
                \edef\@p@sangle{#1} %
}
\def\@p@@ssilent#1{ 
                \@verbosefalse
}
\def\@p@@sprolog#1{\@prologfiletrue\def\@prologfileval{#1}}
\def\@p@@spostlog#1{\@postlogfiletrue\def\@postlogfileval{#1}}
\def\@cs@name#1{\csname #1\endcsname}
\def\@setparms#1=#2,{\@cs@name{@p@@s#1}{#2}}
\def\ps@init@parms{
                \@bbllxfalse \@bbllyfalse
                \@bburxfalse \@bburyfalse
                \@heightfalse \@widthfalse
                \@rheightfalse \@rwidthfalse
                \def\@p@sbbllx{}\def\@p@sbblly{}
                \def\@p@sbburx{}\def\@p@sbbury{}
                \def\@p@sheight{}\def\@p@swidth{}
                \def\@p@srheight{}\def\@p@srwidth{}
                \def\@p@sangle{0}
                \def\@p@sfile{} \def\@p@sbbfile{}
                \def\@p@scost{10}
                \def\@sc{}
                \@prologfilefalse
                \@postlogfilefalse
                \@clipfalse
                \if@noisy
                        \@verbosetrue
                \else
                        \@verbosefalse
                \fi
}
\def\parse@ps@parms#1{
                \@psdo\@psfiga:=#1\do
                   {\expandafter\@setparms\@psfiga,}}
\newif\ifno@bb
\def\bb@missing{
        \if@verbose{
                \ps@typeout{psfig: searching \@p@sbbfile \space  for bounding box}
        }\fi
        \no@bbtrue
        \epsf@getbb{\@p@sbbfile}
        \ifno@bb \else \bb@cull\epsf@llx\epsf@lly\epsf@urx\epsf@ury\fi
}       
\def\bb@cull#1#2#3#4{
        \dimen100=#1 bp\edef\@p@sbbllx{\number\dimen100}
        \dimen100=#2 bp\edef\@p@sbblly{\number\dimen100}
        \dimen100=#3 bp\edef\@p@sbburx{\number\dimen100}
        \dimen100=#4 bp\edef\@p@sbbury{\number\dimen100}
        \no@bbfalse
}
\newdimen\p@intvaluex
\newdimen\p@intvaluey
\def\rotate@#1#2{{\dimen0=#1 sp\dimen1=#2 sp
                  \global\p@intvaluex=\cosine\dimen0
                  \dimen3=\sine\dimen1
                  \global\advance\p@intvaluex by -\dimen3
                  \global\p@intvaluey=\sine\dimen0
                  \dimen3=\cosine\dimen1
                  \global\advance\p@intvaluey by \dimen3
                  }}
\def\compute@bb{
                \no@bbfalse
                \if@bbllx \else \no@bbtrue \fi
                \if@bblly \else \no@bbtrue \fi
                \if@bburx \else \no@bbtrue \fi
                \if@bbury \else \no@bbtrue \fi
                \ifno@bb \bb@missing \fi
                \ifno@bb \ps@typeout{FATAL ERROR: no bb supplied or found}
                        \no-bb-error
                \fi
                \count203=\@p@sbburx
                \count204=\@p@sbbury
                \advance\count203 by -\@p@sbbllx
                \advance\count204 by -\@p@sbblly
                \edef\ps@bbw{\number\count203}
                \edef\ps@bbh{\number\count204}
                \if@angle 
                        \Sine{\@p@sangle}\Cosine{\@p@sangle}
                        {\dimen100=\maxdimen\xdef\r@p@sbbllx{\number\dimen100}
                                            \xdef\r@p@sbblly{\number\dimen100}
                                            \xdef\r@p@sbburx{-\number\dimen100}
                                            \xdef\r@p@sbbury{-\number\dimen100}}
                        \def\minmaxtest{
                           \ifnum\number\p@intvaluex<\r@p@sbbllx
                              \xdef\r@p@sbbllx{\number\p@intvaluex}\fi
                           \ifnum\number\p@intvaluex>\r@p@sbburx
                              \xdef\r@p@sbburx{\number\p@intvaluex}\fi
                           \ifnum\number\p@intvaluey<\r@p@sbblly
                              \xdef\r@p@sbblly{\number\p@intvaluey}\fi
                           \ifnum\number\p@intvaluey>\r@p@sbbury
                              \xdef\r@p@sbbury{\number\p@intvaluey}\fi
                           }
                        \rotate@{\@p@sbbllx}{\@p@sbblly}
                        \minmaxtest
                        \rotate@{\@p@sbbllx}{\@p@sbbury}
                        \minmaxtest
                        \rotate@{\@p@sbburx}{\@p@sbblly}
                        \minmaxtest
                        \rotate@{\@p@sbburx}{\@p@sbbury}
                        \minmaxtest
                        \edef\@p@sbbllx{\r@p@sbbllx}\edef\@p@sbblly{\r@p@sbblly}
                        \edef\@p@sbburx{\r@p@sbburx}\edef\@p@sbbury{\r@p@sbbury}
                \fi
                \count203=\@p@sbburx
                \count204=\@p@sbbury
                \advance\count203 by -\@p@sbbllx
                \advance\count204 by -\@p@sbblly
                \edef\@bbw{\number\count203}
                \edef\@bbh{\number\count204}
}
\def\in@hundreds#1#2#3{\count240=#2 \count241=#3
                     \count100=\count240        %
                     \divide\count100 by \count241
                     \count101=\count100
                     \multiply\count101 by \count241
                     \advance\count240 by -\count101
                     \multiply\count240 by 10
                     \count101=\count240        %
                     \divide\count101 by \count241
                     \count102=\count101
                     \multiply\count102 by \count241
                     \advance\count240 by -\count102
                     \multiply\count240 by 10
                     \count102=\count240        %
                     \divide\count102 by \count241
                     \count200=#1\count205=0
                     \count201=\count200
                        \multiply\count201 by \count100
                        \advance\count205 by \count201
                     \count201=\count200
                        \divide\count201 by 10
                        \multiply\count201 by \count101
                        \advance\count205 by \count201
                     \count201=\count200
                        \divide\count201 by 100
                        \multiply\count201 by \count102
                        \advance\count205 by \count201
                     \edef\@result{\number\count205}
}
\def\compute@wfromh{
                \in@hundreds{\@p@sheight}{\@bbw}{\@bbh}
                \edef\@p@swidth{\@result}
}
\def\compute@hfromw{
                \in@hundreds{\@p@swidth}{\@bbh}{\@bbw}
                \edef\@p@sheight{\@result}
}
\def\compute@handw{
                \if@height 
                        \if@width
                        \else
                                \compute@wfromh
                        \fi
                \else 
                        \if@width
                                \compute@hfromw
                        \else
                                \edef\@p@sheight{\@bbh}
                                \edef\@p@swidth{\@bbw}
                        \fi
                \fi
}
\def\compute@resv{
                \if@rheight \else \edef\@p@srheight{\@p@sheight} \fi
                \if@rwidth \else \edef\@p@srwidth{\@p@swidth} \fi
}
\def\compute@sizes{
        \compute@bb
        \if@scalefirst\if@angle
        \if@width
           \in@hundreds{\@p@swidth}{\@bbw}{\ps@bbw}
           \edef\@p@swidth{\@result}
        \fi
        \if@height
           \in@hundreds{\@p@sheight}{\@bbh}{\ps@bbh}
           \edef\@p@sheight{\@result}
        \fi
        \fi\fi
        \compute@handw
        \compute@resv}

\def\psfig#1{\vbox {
        \ps@init@parms
        \parse@ps@parms{#1}
        \compute@sizes
        \ifnum\@p@scost<\@psdraft{
                \special{ps::[begin]    \@p@swidth \space \@p@sheight \space
                                \@p@sbbllx \space \@p@sbblly \space
                                \@p@sbburx \space \@p@sbbury \space
                                startTexFig \space }
                \if@angle
                        \special {ps:: \@p@sangle \space rotate \space} 
                \fi
                \if@clip{
                        \if@verbose{
                                \ps@typeout{(clip)}
                        }\fi
                        \special{ps:: doclip \space }
                }\fi
                \if@prologfile
                    \special{ps: plotfile \@prologfileval \space } \fi
                \if@decmpr{
                        \if@verbose{
                                \ps@typeout{psfig: including \@p@sfile.Z \space }
                        }\fi
                        \special{ps: plotfile "`zcat \@p@sfile.Z" \space }
                }\else{
                        \if@verbose{
                                \ps@typeout{psfig: including \@p@sfile \space }
                        }\fi
                        \special{ps: plotfile \@p@sfile \space }
                }\fi
                \if@postlogfile
                    \special{ps: plotfile \@postlogfileval \space } \fi
                \special{ps::[end] endTexFig \space }
                \vbox to \@p@srheight sp{
                        \hbox to \@p@srwidth sp{
                                \hss
                        }
                \vss
                }
        }\else{
                \if@draftbox{           
                        \hbox{\frame{\vbox to \@p@srheight sp{
                        \vss
                        \hbox to \@p@srwidth sp{ \hss \@p@sfile \hss }
                        \vss
                        }}}
                }\else{
                        \vbox to \@p@srheight sp{
                        \vss
                        \hbox to \@p@srwidth sp{\hss}
                        \vss
                        }
                }\fi

        }\fi
}}
\psfigRestoreAt
\let\@=\LaTeXAtSign

\makeatletter%
\def\nottoobig#1{{\hbox{$\left#1\vcenter to1.111\ht\strutbox{}\right.\n@space$}}}
\makeatother%

\topsep 8pt plus2pt minus4pt   %

\makeatletter%
\def\@begintheorem#1#2{\trivlist\item[\hskip\labelsep{\bf #1\ #2}]}
\makeatother
\makeatletter %
\newcommand{\C}[1]{ {\rm {#1}} }
\newcommand{\band}{\bigwedge}
\newcommand{\Band}[3]{(\bigwedge#1\!\!:\,#2\!\!:\,#3)}
\newcommand{\bor}{\bigvee}
\newcommand{\Bor}[3]{(\bigvee#1\!\!:\,#2\!\!:\,#3)}
\newcommand{\Forall}[3]{(\forall #1\!\!:\,#2\!\!:\,#3)}
\newcommand{\Exists}[3]{(\exists #1\!\!:\,#2\!\!:\,#3)}
\newcommand{\Union}[3]{(\bigcup #1\!\!:\,#2\!\!:\,#3)}
\def\union{\,\bigcup\limits\,}
\newcommand{\true}{\mbox{\it true}}
\newcommand{\false}{\mbox{\it false}}
\newcommand{\SUM}[3]{ (\sum #1 \!\! : \, #2 \!\!:\, #3) }
\newcommand{\IFS}{\mbox{\bf if}}
\newcommand{\IF}[1]{ \mbox{\bf if} \, #1 \, \rightarrow \, }
\newcommand{\GC}[2]{ #1 \, \rightarrow \, #2 }
\newlength{\filength}
\settowidth{\filength}{\mbox{\bf f{}i}}
\newsavebox{\gcbox}
\sbox{\gcbox}{\framebox[\filength]{\rule{0ex}{2ex}}}
\newcommand{\BB}[1]{\usebox{\gcbox}\; #1 \, \rightarrow \, }
\newcommand{\FI}{\; \mbox{\bf f{}i}}
\newcommand{\Skip}{ \mbox{\bf skip} }
\newcommand{\DOS}{\mbox{\bf do}}
\newcommand{\DO}[1]{\mbox{\bf do} \, #1 \, \rightarrow \,}
\newcommand{\OD}{\mbox{\bf od}}
\newcommand{\cobegin}{{\bf cobegin}\,}
\renewcommand{\|}{\, //  \,}
\newcommand{\coend}{\,{\bf coend}}
\newcommand{\Set}[1]{ \hbox{\bf\{} #1 \hbox{\bf\}}}
\newcommand{\Bag}[1]{ \{\!| #1  |\!\}}
\newlength{\leftjustindent}
\newlength{\@leftjustindent}
\setlength{\@leftjustindent}{\leftmargin}
\def\leftjust{\let\\\@leftjustcr\let\end\@endleftjust
  \addtolength{\@leftjustindent}{\leftjustindent}
  \vcenter\bgroup
  \halign\bgroup
    \hbox to\displaywidth{
      \rule{\@leftjustindent}{0ex}$\displaystyle##$\hfill
      }\crcr
}
\def\endleftjust{\crcr\egroup\egroup\endgroup}
\def\@endleftjust#1{\crcr\egroup\egroup\@checkend{#1}\endgroup}
\def\@leftjustcr{\crcr}

\newcommand{\hoare}[3]{\{{#1}\}\:{#2}\:\{{#3}\}}
\renewcommand{\wp}[2]{ {\it wp}({#1},{#2})}
\newcommand{\assert}[1]{\!\{#1\}}
\newcommand{\atom}[1]{\langle\,{#1}\,\rangle}
\newcommand{\lbl}[1]{{#1 \!:\;\,}}
\newcommand{\pre}[1]{ {\it pre}({#1})}
\newcommand{\post}[1]{ {\it post}({#1}) }
\newcommand{\NI}[2]{ {\it NI}({#1},{#2}) }
\newcommand{\equi}[3]{ {  {\rm E}_{#1}^{#2}({#3}) }    }
\newcommand{\red}[3]{ {  {\rm R}_{#1}^{#2}({#3}) }    }
\newcommand{\sparse}{{{\rm SPARSE}}}
\newcommand{\tally}{{{\rm TALLY}}}
\newcommand{\inferfrom}[2]{\begin{array}[t]{c}\displaystyle
   \frac{#1}{#2}\end{array}}
\newtheorem{theorem}{Theorem}[section]

\newtheorem{corollary}[theorem]{Corollary}

\newcommand{\qedblob}{\mbox{\rule[-1.5pt]{5pt}{10.5pt}}}
\def\literalqed{{\ \nolinebreak\hfill\mbox{\qedblob\quad}}}
\def\qedcareful{\literalqed}
\def\qed{\literalqed}
\def\trueloveqed{{\ \nolinebreak\hfill\mbox{\boldmath
\Huge$ \Box$}\nolinebreak\mbox{$\!\!\!\!\!\!
{}^{\normalsize\heartsuit}$}}}

\newcommand{\singlespacing}{\let\CS=
\@currsize\renewcommand{\baselinestretch}{1}\tiny\CS}
\newcommand{\singlespacingplus}{\let\CS=
\@currsize\renewcommand{\baselinestretch}{1.25}\tiny\CS}
\newcommand{\doublespacing}{\let\CS=
\@currsize\renewcommand{\baselinestretch}{1.75}\tiny\CS}
\newcommand{\draftspacing}{\let\CS=
\@currsize\renewcommand{\baselinestretch}{1.65}\tiny\CS}
\newcommand{\normalspacing}{\singlespacing}
\makeatother%

\hyphenation{theory area areas theorem theorems par-allel par-allelize par-allelized threshold Hemaspaan-dra}

\newtheorem{definition}[theorem]{Definition}

\flushbottom{}
\makeatletter
\clubpenalty=\@highpenalty
\widowpenalty=\@highpenalty
\makeatother

\let\BLS=\baselinestretch
\lefthyphenmin=100
\emergencystretch=2em

\makeatletter
\newcommand{\niceonespacing}{\let\CS=\@currsize\renewcommand{\baselinestretch}{1.1}\tiny\CS}\newcommand{\nicetwospacing}{\let\CS=\@currsize\renewcommand{\baselinestretch}{1.2}\tiny\CS}
\newcommand{\nicethreespacing}{\let\CS=\@currsize\renewcommand{\baselinestretch}{1.3}\tiny\CS}
\newcommand{\singlespacingplusplus}{\let\CS=\@currsize\renewcommand{\baselinestretch}{1.35}\tiny\CS}
\newcommand{\nicefivespacing}{\let\CS=\@currsize\renewcommand{\baselinestretch}{1.5}\tiny\CS}
\newcommand{\nicesixspacing}{\let\CS=\@currsize\renewcommand{\baselinestretch}{1.6}\tiny\CS}
\newcommand{\nicefoospacing}{\let\CS=\@currsize\renewcommand{\baselinestretch}{1.05}\tiny\CS}
\makeatother

\makeatletter
\def\@cite#1#2{[#1\if@tempswa , #2\fi]}
\makeatother

\makeatletter
\def\@citex[#1]#2{\if@filesw\immediate\write\@auxout{\string\citation{#2}}\fi
  \def\@citea{}\@cite{\@for\@citeb:=#2\do
    {\@citea\def\@citea{,\linebreak[0]}\@ifundefined
       {b@\@citeb}{{\bf ?}\@warning
       {Citation `\@citeb' on page \thepage \space undefined}}%
\hbox{\csname b@\@citeb\endcsname}}}{#1}}
\makeatother

\newcommand{\sharpp}{{\rm \#P}}
\newcommand{\sharpsat}{{\rm \#SAT}}
\newcommand{\sat}{{\rm SAT}}
\newcommand{\qbf}{{\rm QBF}}
\newcommand{\parityp}{{\rm \oplus P}}
\newcommand{\up}{{\rm UP}}
\newcommand{\us}{{\rm US}}
\newcommand{\fewnp}{{\rm FewNP}}
\newcommand{\fewp}{{\rm FewP}}
\newcommand{\coup}{{\rm coUP}}
\newcommand{\e}{{\rm E}}
\renewcommand{\exp}{{\rm EXP}}
\newcommand{\NE}{{\rm NE}}
\renewcommand{\ne}{{\rm NE}}
\newcommand{\nexp}{{\rm NEXP}}
\newcommand{\p}{{\rm P}}
\newcommand{\littlep}{{\rm p}}
\newcommand{\NP}{{\rm NP}}
\newcommand{\npnp}{{\rm NP^{NP}}}
\newcommand{\bh}{{\rm BH}}
\newcommand{\BH}{{\rm BH}}
\newcommand{\BPP}{{\rm BPP}}
\newcommand{\Prob}{{\rm Prob}}
\newcommand{\MOD}{{\rm MOD}}
\newcommand{\BPTIME}{{\rm BPTIME}}
\newcommand{\ZPTIME}{{\rm ZPTIME}}
\newcommand{\DTIME}{{\rm DTIME}}
\newcommand{\dtime}{{\rm DTIME}}
\newcommand{\BPSPACE}{{\rm BPSPACE}}

\newcommand{\ie}{{\mbox{i.e.}}}
\newcommand{\inter}{{\cap}}
\newcommand{\spp}{{\rm SPP}}
\newcommand{\gapp}{{\rm GapP}}
\newcommand{\pl}{{\rm PL}}
\def\sstar{\Sigma^{*}}

\newcommand{\np}{{\rm NP}}
\newcommand{\nt}{{\rm NT}}
\newcommand{\nnt}{{\rm NNT}}
\newcommand{\parityoptp}{{\rm \oplus{}OptP}}
\newcommand{\optp}{{\rm OptP}}
\newcommand{\diffp}{{\rm D^P}}
\newcommand{\pp}{{\rm PP}}
\newcommand{\bpp}{{\rm BPP}}
\newcommand{\zpp}{{\rm ZPP}}
\newcommand{\cor}{{\rm coR}}
\newcommand{\npc}{$\np$-com\-plete}
\newcommand{\conp}{{\rm coNP}}
\newcommand{\pspace}{{\rm PSPACE}}
\newcommand{\eespace}{{\rm EESPACE}}
\newcommand{\dspace}{{\rm DSPACE}}
\newcommand{\psp}{{\pspace}}
\newcommand{\pnexp}{{\p^\nexp}}
\newcommand{\npnexp}{{\np^\nexp}}
\newcommand{\nenp}{{\ne^\np}}
\newcommand{\enp}{{\e^\np}}
\newcommand{\pnp}{{\p^\np}}
\newcommand{\pnplog}{{\p^{\np[\log ]}}}
\newcommand{\pij}{{\p^{\bh_i:\bh_j}}}
\newcommand{\pji}{{\p^{\bh_j:\bh_i}}}
\newcommand{\nexpnp}{{\nexp^\np}}
\newcommand{\coNP}{{\rm coNP}}
\newcommand{\cone}{{\rm CONE}}
\newcommand{\sigmatwozero}{{\Sigma_2^0}}
\newcommand{\pitwozero}{{\Pi_2^0}}
\newcommand{\pithreezero}{{\Pi_3^0}}
\newcommand{\sigmathreezero}{{\Sigma_3^0}}
\newcommand{\sigmatwo}{{\Sigma_2^{\littlep}}}
\newcommand{\sigmathree}{{\Sigma_3^{\littlep}}}
\newcommand{\sigmafour}{{\Sigma_4^{\littlep}}}
\newcommand{\sigmafive}{{\Sigma_5^{\littlep}}}
\newcommand{\sigmak}{{\Sigma_k^{\littlep}}}
\newcommand{\sigmai}{{\Sigma_i^{\littlep}}}
\newcommand{\sigmaj}{{\Sigma_j^{\littlep}}}
\newcommand{\pitwo}{{\Pi_2^{\littlep}}}
\newcommand{\pithree}{{\Pi_3^{\littlep}}}
\newcommand{\pifour}{{\Pi_4^{\littlep}}}
\newcommand{\pifive}{{\Pi_5^{\littlep}}}
\newcommand{\thetatwo}{{\Theta_2^{\littlep}}}
\newcommand{\deltatwo}{{\Delta_2^{\littlep}}}
\newcommand{\poly}{{\rm poly}}
\newcommand{\ph}{{\rm PH}}
\newcommand{\few}{{\rm Few}}
\newcommand{\fewch}{{\rm FewCH}}
\newcommand{\eh}{{\rm EH}}
\def\bull{\vrule height .9ex width .8ex depth -.1ex }
\newcommand{\blob}{\mbox{\rule[-1.5pt]{5pt}{10.5pt}}}
\newcommand{\lindent}{\qquad}
\newcommand{\magicnum}{{ n^{\frac{1-\epsilon}{\epsilon}+\delta}}}
\newcommand{\fsup}{{\,f_{super}\,}}
\newcommand{\fred}{{\,f_{reduced}\,}}
\newcommand{\pne}{{\p^\ne}}
\newcommand{\npne}{{\np^\ne}}
\newcommand{\nnexarg}{{\nxx^\nexx (x) }}
\newcommand{\nnexx}{{\nxx^\nexx  }}
\newcommand{\nnex}{{\nxx^\nexx }}
\newcommand{\expnp}{{\exp^\np }}
\newcommand{\nxx}{{\rm N_{17}}}
\newcommand{\nexx}{{\rm NE_{21}}}
\newcommand{\seh}{{\rm SEH}}
\newcommand{\sexph}{{\rm SEXPH}}
\newcommand{\pstar}{{\p_\star}}
\newcommand{\nestar}{{\ne_{\,\star}}}
\newcommand{\supersetproper}{  \stackrel{\scriptscriptstyle\superset}{\scriptscriptstyle\not-}}
\newcommand{\subsetproper}{  \stackrel{\scriptscriptstyle\subset}{\scriptscriptstyle\not-}}
\newcommand{\superset}{\supset}
\newcommand{\superseteq}{\supseteq}

\newcommand{\substar}{\mbox{$\subset^*$}}
\newcommand{\superstar}{\mbox{$\superset^*$}}

\def\unionfromc{\,\textstyle\bigcup_{\scriptstyle c}\,}
\def\unionfromk{\,\textstyle\bigcup_{\scriptstyle k}\,}

\newcommand{\co}{{\rm co}}
\newcommand{\thetathree}{{\Theta_3^{\littlep}}}
\newcommand{\deltathree}{{\Delta_3^{\littlep}}}
\newcommand{\sigmajmone}{{{\rm \Sigma}_j-1^{\littlep}}}
\newcommand{\sigmakminusone}{{{\rm \Sigma}_{k-1}^{\littlep}}}
\newcommand{\sigmakminustwo}{{{\rm \Sigma}_{k-2}^{\littlep}}}
\newcommand{\sigmakplusone}{{\rm \Sigma}_{k+1}^{\littlep}}
\newcommand{\sigmakmone}{{\rm \Sigma}_{k-1}^{\littlep}}
\newcommand{\sigmakmtwo}{{\rm \Sigma}_{k-2}^{\littlep}}
\newcommand{\sigmakpone}{{\rm \Sigma}_{k+1}^{\littlep}}
\newcommand{\sigmakptwo}{{\rm \Sigma}_{k+2}^{\littlep}}
\newcommand{\pik}{{\rm \Pi}_{k}^{\littlep}}
\newcommand{\pikpone}{{\rm \Pi}_{k+1}^{\littlep}}
\newcommand{\sigmaipone}{{\rm \Sigma}_{i+1}^{\littlep}}
\newcommand{\sigmaiptwo}{{\rm \Sigma}_{i+2}^{\littlep}}
\newcommand{\sigmaimone}{{\rm \Sigma}_{i-1}^{\littlep}}
\newcommand{\thetak}{{\Theta_k^{\littlep}}}
\newcommand{\deltak}{{\Delta_k^{\littlep}}}
\newcommand{\deltai}{{\Delta_i^{\littlep}}}
\newcommand{\psigkone}{ {\p^{\sigmak[1]}}}
\newcommand{\psigktwo}{ {\p^{\sigmak[2]}}}
\newcommand{\psigkjqueries}{ {\p^{\sigmak[j]}}}
\newcommand{\psigkjplusone}{ {\p^{\sigmak[j+1]}}}
\newcommand{\psigkmqueries}{ {\p^{\sigmak[m]}}}
\newcommand{\psigkmplusone}{ {\p^{\sigmak[m+1]}}}
\newcommand{\pnpbigolog}{{\p^{\np[{\cal O}(\log)]}}}
\newcommand{\pnpbigoone}{{\p^{\np[{\cal O}(1)]}}}

\newcommand{\wh}[1]{\widehat{#1}}
\newcommand{\lp}{ {L_{\p}}}
\newcommand{\lpone}{ {L_{\p^{\p[1]}}}}
\newcommand{\lnpone}{ {L_{\p^{\np[1]}}}}
\newcommand{\lnp}{ {L_{\np}}}
\newcommand{\lsigk}{ {L_{\sigmak}}}
\newcommand{\lpik}{ {L_{\pik}}}
\newcommand{\lsigtwo}{ {L_{\sigmatwo}}}
\newcommand{\lsigthree}{ {L_{\sigmathree}}}
\newcommand{\lsigfour}{ {L_{\sigmafour}}}
\newcommand{\lsigfive}{ {L_{\sigmafive}}}
\newcommand{\lsigtwobar}{ \overline{L_{\sigmatwo}}}
\newcommand{\lsigthreebar}{ \overline{L_{\sigmathree}}}
\newcommand{\lsigfourbar}{ \overline{L_{\sigmafour}}}
\newcommand{\lsigfivebar}{ \overline{L_{\sigmafive}}}
\newcommand{\lsigkprime}{ {L'_{\sigmak}}}
\newcommand{\lsigi}{ {L_{\sigmai}}}
\newcommand{\lsigj}{ {L_{\sigmaj}}}
\newcommand{\lsigipone}{ {L_{\sigmaipone}}}
\newcommand{\lsigiptwo}{ {L_{\sigmaiptwo}}}
\newcommand{\lsigkpone}{ {L_{\sigmakpone}}}
\newcommand{\lsigkptwo}{ {L_{\sigmakptwo}}}
\newcommand{\lsigimone}{ {L_{\sigmaimone}}}
\newcommand{\lsigkmone}{ {L_{\sigmakmone}}}
\newcommand{\lsigkmtwo}{ {L_{\sigmakmtwo}}}
\newcommand{\lsigkmtwoprime}{ {L'_{\sigmakmtwo}}}
\newcommand{\lsigkmtwoprimeprime}{ {L''_{\sigmakmtwo}}}
\newcommand{\ldeli}{ {L_{\deltai}}}
\newcommand{\ldelkmone}{ {L_{\deltakmone}}}
\newcommand{\psigthreeone}{  {\p^{\sigmathree[1]}}}
\newcommand{\psigthreetwo}{  {\p^{\sigmathree[2]}}}
\newcommand{\psigthreetwott}{  {\p^{\sigmathree}_{2\hbox{-}{\rm tt}}}}
\newcommand{\psigthreethree}{  {\p^{\sigmathree[3]}}}
\newcommand{\psigthreethreett}{  {\p^{\sigmathree}_{3\hbox{-}{\rm tt}}}}
\newcommand{\psigtwoone}{  {\p^{\sigmatwo[1]}}}
\newcommand{\psigtwotwo}{  {\p^{\sigmatwo[2]}}}
\newcommand{\psigtwotwott}{  {\p^{\sigmatwo}_{2\hbox{-}{\rm tt}}}}
\newcommand{\psigtwothree}{  {\p^{\sigmatwo[3]}}}
\newcommand{\psigtwothreett}{  {\p^{\sigmatwo}_{3\hbox{-}{\rm tt}}}}
\newcommand{\psigkmtt}{  {\p^{\sigmak}_{m \hbox{-}{\rm tt}}}}
\newcommand{\psigkmponett}{  {\p^{\sigmak}_{(m+1) \hbox{-}{\rm tt}}}}
\newcommand{\psigimk}{  {\p^{(\sigmai , \sigmak)}_{1,m \hbox{-}{\rm tt}}}}
\newcommand{\psigjustnpmk}{  {\p^{(\np , \sigmak)}_{1,m \hbox{-}{\rm tt}}}}
\newcommand{\psigjmk}{  {\p^{(\sigmaj , \sigmak)}_{1,m \hbox{-}{\rm tt}}}}
\newcommand{\psigik}{  {\p^{(\sigmai , \sigmak)}}}
\newcommand{\psigjustnpk}{  {\p^{(\np , \sigmak)}}}
\newcommand{\psigij}{  {\p^{(\sigmai , \sigmaj)}}}
\newcommand{\psigkponek}{  {\p^{(\Sigma_{k+1}^{\littlep} : \sigmak)}}}
\newcommand{\psigione}{ {\p^{\sigmai[1]}}}
\newcommand{\psigipone}{{\p^{\Sigma_{i+1}^{\littlep}[1]}}}
\newcommand{\psigjone}{ {\p^{\sigmaj[1]}}}
\newcommand{\psigkk}{ {\p^{\sigmak : \sigmak}}}
\newcommand{\lpsigi}{ {L_{\psigione}}}
\newcommand{\lpsigj}{ {L_{\psigjone}}}
\newcommand{\lpsigipone}{{L_{\p^{\Sigma_{i+1}^{\littlep}[1]}}}}
\newcommand{\diffmsigk}{{\rm DIFF}_m(\sigmak)}
\newcommand{\diffmpik}{{\rm DIFF}_m(\pik)}
\newcommand{\dsigtwo}{{\rm D}\!\cdot\!\sigmatwo}
\newcommand{\dsigthree}{{\rm D}\!\cdot\!\sigmathree}
\newcommand{\diffssigi}{{\rm DIFF}_s(\sigmai)}
\newcommand{\diffspii}{{\rm DIFF}_s(\pii)}
\newcommand{\diffsplusonesigi}{{\rm DIFF}_{s+1}(\sigmai)}
\newcommand{\diffmplusonesigk}{{\rm DIFF}_{m+1}(\sigmak)}
\newcommand{\diff}{{\rm DIFF}}
\newcommand{\codiffmsigk}{{\rm co}{\diffmsigk}}
\newcommand{\codsigtwo}{{\rm co}{\dsigtwo}}
\newcommand{\codsigthree}{{\rm co}{\dsigthree}}
\newcommand{\codiffmplusonesigk}{{\rm co}{\diffmplusonesigk}}
\newcommand{\ldiffmsigk}{{L_{\diffmsigk}}}
\newcommand{\ldiffmpik}{{L_{\diffmpik}}}
\newcommand{\ldsigtwo}{{L_{ {\rm D}\cdot\sigmatwo}}}
\newcommand{\lprimedsigtwo}{{L'_{ {\rm D}\cdot\sigmatwo}}}
\newcommand{\ldsigtwobar}{\overline{\ldsigtwo}}
\newcommand{\ldsigthree}{{L_{{\rm D}\cdot\sigmathree}}}
\newcommand{\ldsigthreebar}{\overline{\ldsigthree}}
\newcommand{\ldiffssigi}{{L_{\diffssigi}}}
\newcommand{\lhatdiffmsigk}{{\wh{L}_{\diffmsigk}}}
\newcommand{\deltatilde}{\tilde{\Delta}}
\newcommand{\bolddelta}{\mbox{\boldmath$\Delta$\unboldmath}}

\newcommand{\lbar}{{\overline{L}}}
\newcommand{\proof}{{\bf Proof:}\quad}
\newcommand{\natnum}{\mathbb{N}}
\newcommand{\dlangle}{\mbox{$\langle\hspace{-.1cm}\langle$}}
\newcommand{\drangle}{\mbox{$\rangle\hspace{-.1cm}\rangle$}}
\newcommand{\newlozenge}{\setlength{\fboxsep}{0pt}\setlength{\fboxrule}{.7pt}\framebox[6pt]{\rule{0pt}{9pt}}}

\singlespacing

\def\pair#1{{{\langle\!\!~#1~\!\!\rangle}}}
\def\pairs#1{{{\langle\!\!~#1~\!\!\rangle}}}
\newcommand{\piso}{\mbox{$\littlep$-iso\-mor\-phic}}
\newcommand{\manyonea}{\mbox{$\,\leq_{\rm m}^{{\littlep},\,A}\,$}}
\newcommand{\manyone}{\mbox{$\,\leq_{\rm m}^{{\littlep}}$\,}}
\newcommand{\Turing}{\mbox{$\,\leq_{\rm T}^{{\littlep}}$\,}}
\newcommand{\paiso}{\mbox{$\littlep^A$-iso\-mor\-phic}}
\newcommand{\pisoa}{\paiso}
\newcommand{\pisoam}{\mbox{$\littlep^A$-iso\-mor\-phism}}
\newcommand{\pisom}{\mbox{$\littlep$-iso\-mor\-phism}}
\newcommand{\pselective}{\mbox{$\p$-selec\-tive}}
\newcommand{\sigmastar}{\mbox{$\Sigma^\ast$}}
\newcommand{\pisnp}{\mbox{$\p=\np$}}
\newcommand{\usuba}{\mbox{$U_A$}}
\newcommand{\univsuba}{\mbox{$Univ_A$}}
\newcommand{\pisnotnp}{\mbox{$\p\neq\np$}}
\newcommand{\lb}{\mbox{\{}}
\newcommand{\rb}{\mbox{\}}}
\newcommand{\pa}{\mbox{$\p^A$}}
\newcommand{\calf}{\mbox{$\cal F$}}
\newcommand{\calc}{\mbox{$\cal C$}}
\newcommand{\cald}{\mbox{$\cal D$}}
\newcommand{\calcone}{{\cal C}_1}
\newcommand{\calctwo}{{\cal C}_2}
\newcommand{\npa}{\mbox{$\np^A$}}
\newcommand{\conpa}{\mbox{$\conp^A$}}
\newcommand{\upa}{\mbox{$\up^A$}}
\newcommand{\sparses}{\mbox{ sparse $S\,$}}
\newcommand{\bigo}{\mbox{$\cal O$}}
\newcommand{\condition}{\,\nottoobig{|}\:}
\def\land{{\; \wedge \;}}

\newcommand{\parallelnp}{\mbox{$\p_{||}^{\np}$}}
\newcommand{\rp}{\rm R}
\newcommand{\corp}{{\rm coR}}
\newcommand{\ceqp}{{\rm C_{\!=\!}P }}
\newcommand{\pclose}{\rm P-close}
\newcommand{\apt}{\rm APT}
\newcommand{\ppoly}{\rm P/poly}
\newcommand{\dr}{\mbox{\tt Carroll Ranking}}
\newcommand{\dw}{\mbox{\tt Carroll Winner}}
\newcommand{\ds}{\mbox{\tt Carroll Score}}
\newcommand{\mee}{\mbox{\tt MEE}}
\sloppy

\begin{document}

\bibliographystyle{alpha}

\title{What's Up with Downward Collapse: Using the Easy-Hard Technique to Link Boolean and Polynomial Hierarchy Collapses\thanks{Supported in part 
by grants 
NSF-CCR-9322513 and 
NSF-INT-9513368/\protect\linebreak[0]DAAD-315-PRO-fo-ab.}}

\date{Technical Report UR-CS-TR-98-682 \\ February, 1998}

\author{{\em Edith Hemaspaandra\/}\footnote{
\protect\singlespacing
{\tt edith@bamboo.lemoyne.edu}.
Work done in part while 
visiting Friedrich-Schiller-Universit\"at Jena.}
\\
Department of Mathematics \\
Le Moyne College \\
Syracuse, NY 13214 \\ 
USA
\and
{\em Lane A. Hemaspaandra\/}\footnote{
\protect\singlespacing
{\tt lane@cs.rochester.edu}.
Work done in part while 
visiting Friedrich-Schiller-Universit\"at Jena.} 
\\
Department of Computer Science \\
University of Rochester \\
Rochester, NY 14627 \\
USA
\and
{\em Harald Hempel\/}\footnote{
\protect\singlespacing
{\tt hempel@informatik.uni-jena.de}.
Work done in part while 
visiting Le Moyne College.}
\\
Institut f\"ur Informatik \\
Friedrich-Schiller-Universit\"at Jena \\
07740 Jena
\\
Germany 
}

{\singlespacing

\maketitle

}

{

\singlespacing 

\noindent{\bf Abstract:} \quad
During the past decade, nine papers have obtained 
increasingly strong 
consequences from the assumption that boolean or bounded-query 
hierarchies collapse.
The final four papers of this nine-paper progression actually 
achieve downward collapse---that is, they show that high-level 
collapses induce collapses at (what beforehand were thought to be) 
lower complexity levels.
For example, for each $k\geq 2$ it is now 
known that if $\psigkone=\psigktwo$ 
then $\ph=\sigmak$.
This article surveys the history, the results, and the technique---the 
so-called easy-hard method---of these nine papers.
\begin{enumerate}
\item J.~Kadin. The polynomial time hierarchy collapses if the boolean 
hierarchy collapses. 
{\it SIAM Journal on Computing,}
17(6):1263-1282, 1988.
Erratum appears in the same journal, 20(2):404.
\item K.~Wagner. Number-of-query hierarchies.
Technical Report 158, Universit\"at Augsburg,
Institut f\"ur Mathematik, Augsburg, Germany, October 1987.
\item K.~Wagner. Number-of-query hierarchies.
Technical Report 4, Universit\"at W\"urzburg,
Institut f\"ur Informatik, W\"urzburg, Germany, February 1989.
\item R.~Chang and J.~Kadin. The boolean hierarchy and the polynomial 
hierarchy: A closer connection.
{\it SIAM Journal on Computing,}
25(2):340-354, 1996.
\item R.~Beigel, R.~Chang, and M.~Ogiwara. A relationship between 
difference hierarchies and relativized polynomial hierarchies.
{\it Mathematical Systems Theory,}
26(3):293-310, 1993.
\item E.~Hemaspaandra, L.~Hemaspaandra, and H.~Hempel. An upward separation in the polynomial hierarchy.
Technical Report Math/Inf/96/15, Friedrich-Schiller-Universit\"at Jena,
Fakult\"at f\"ur Mathematik und Informatik,
Jena, Germany, June 1996.
\item E.~Hemaspaandra, L.~Hemaspaandra, and H.~Hempel. A downward collapse 
within the polynomial hierarchy. 
{\it SIAM Journal on Computing.}
To appear.
\item H.~Buhrman and L.~Fortnow. Two queries.
Proceedings of the 13th Annual IEEE Conference on
Computational Complexity.  To appear.

\item E.~Hemaspaandra, L.~Hemaspaandra, and H.~Hempel. Translating 
equality downwards. 
Technical Report TR-657, University of Rochester, 
Department of Computer Science, Rochester, NY, April 1997.
\end{enumerate}

} %

\section{Introduction}
\label{s:intro}

Does the polynomial hierarchy collapse if the 
boolean hierarchy or the bounded-query hierarchy collapse?
Kadin~\cite{kad:joutdatedbychangkadin:bh} was able to answer this question 
affirmatively with the help of the easy-hard technique.
Until 1995 this technique (slightly modified) has been used five times to 
obtain stronger and stronger collapses of the polynomial hierarchy from the 
assumption that the boolean hierarchy (also called difference hierarchy) or 
the bounded-query hierarchy collapse. 
All those results have in common that a collapse of the 
boolean or bounded-query hierarchy induces a collapse 
of the polynomial hierarchy at a higher level.
In 1996 Hemaspaandra, Hemaspaandra, and 
Hempel~\cite{hem-hem-hem:jtoappear:downward-translation} and 
also Buhrman and Fortnow~\cite{buh-for:t:two-queries} obtained  so 
called downward collapse results within the polynomial hierarchy, 
the easy-hard technique playing a crucial role in the proofs.
The term downward collapse refers to the fact that the collapse of larger 
classes implies the collapse of smaller classes,  
the {\em collapse\/} translates {\em downwards\/}. 
It seems that the time is right to take a close look at the easy-hard method,  
especially its applications for collapsing the polynomial hierarchy.

This survey will be structured as follows. 
The timeline of the nine papers using the 
easy-hard method is given in Section~\ref{s:timeline}. 
Section~\ref{s:results} lists the key results of these nine 
papers. 
In Section~\ref{s:improved} we prove a new 
and stronger result regarding to what exact level 
the polynomial hierarchy collapses if the boolean hierarchy over $\sigmak$ 
collapses.
Section~\ref{s:evolution} gives an overview of the history of the easy-hard 
technique. 
In particular, we will first informally discuss the contributions 
of the various papers to the evolution of the easy-hard technique.
Second, will rigorously prove a special case of the main theorem  
of each of the papers and so illustrate the development of the 
easy-hard technique to yield stronger and stronger results.
Finally, Section~\ref{s:questions} suggests interesting open issues 
related to the topic of this article.

\section{The Timeline}
\label{s:timeline}

Table~\ref{tab:timeline} gives the relevant dates and citations for all 
nine papers, in particular the dates of the earliest versions and pointers 
to the earliest and most recent versions.

\begin{table}[t]
\begin{tabular}{|l|c|c|c|}\hline
&Date of&&\\
\qquad\qquad \raisebox{1.5ex}[-1.5ex]{Author(s)}&Earliest Version&
\raisebox{1.5ex}[-1.5ex]{Earliest Version}&
\raisebox{1.5ex}[-1.5ex]{Most Recent Version}\\ \hline \hline
Kadin&6/87&\cite{kad:toutdate:bh}&\cite{kad:joutdatedbychangkadin:bh}\\ 
\hline
Wagner&10/87&\cite{wag:t:n-o-q-87version}&\cite{wag:t:n-o-q-87version}\\ \hline
Wagner&2/89&\cite{wag:t:n-o-q-89version}&\cite{wag:t:n-o-q-89version}\\ \hline
Chang/Kadin&5/89&\cite{cha-kad:toutdated:bh}&
\cite{cha-kad:j:closer}\\ \hline
Beigel/Chang/Ogiwara&1/91&\cite{bei-cha-ogi:tOUT:boolean}&
\cite{bei-cha-ogi:j:difference-hierarchies}\\ \hline
Hemaspaandra/Hemaspaandra/&&&\\
Hempel&\raisebox{1.5ex}[-1.5ex]{6/96}&
\raisebox{1.5ex}[-1.5ex]{\cite{hem-hem-hem:tOutBydown-sep:upward-sep}}&
\raisebox{1.5ex}[-1.5ex]{\cite{hem-hem-hem:tOutBydown-sep:upward-sep}}\\
\hline
Hemaspaandra/Hemaspaandra/&&&\\
Hempel&\raisebox{1.5ex}[-1.5ex]{7/96}&
\raisebox{1.5ex}[-1.5ex]
{\cite{hem-hem-hem:tOutByConf:downward-translation}}&
\raisebox{1.5ex}[-1.5ex]{\cite{hem-hem-hem:jtoappear:downward-translation}}\\
\hline
Buhrman/Fortnow&9/96&\cite{buh-for:t:two-queries}&
\cite{buh-for:ctoappear:two-queries}\\ \hline
Hemaspaandra/Hemaspaandra/&&&\\
Hempel&\raisebox{1.5ex}[-1.5ex]{4/97}&
\raisebox{1.5ex}[-1.5ex]{\cite{hem-hem-hem:t:translating-downwards}}&
\raisebox{1.5ex}[-1.5ex]{\cite{hem-hem-hem:t:translating-downwards}}\\ 
\hline
\end{tabular}
\caption{\label{tab:timeline}Timeline.}
\end{table}

\section{The Results}
\label{s:results}

The pace and amount of change underwent by the easy-hard technique, 
in particular how it has been used to collapse the polynomial hierarchy, 
can best be 
seen by a close look at the key results of the relevant papers.
In this section we will list their main theorems (or most 
charismatic results).
This will also nicely illustrate the improvements
each made with respect to 
the results that were known before it.

A definition of the basic concepts involved seems appropriate at this point.
The polynomial hierarchy was introduced by Stockmeyer~\cite{sto:j:poly}.

\begin{definition}
\begin{enumerate}
\item For any set of languages $\calc$, let 
$\co\calc=\{\overline L\condition L \in \calc\}$.
\item\cite{sto:j:poly}\nopagebreak%
\begin{enumerate}%
\item $\Delta_0^{\littlep}=\Sigma_0^{\littlep}=\Pi_0^{\littlep}=\p$.
\item For all $k\geq 0$, $\Delta_{k+1}^{\littlep}=\p^{\sigmak}$, 
$\Sigma_{k+1}^{\littlep}=\np^{\sigmak}$, and 
$\Pi_{k+1}^{\littlep}=\co \Sigma_{k+1}^{\littlep}$. 
\item The polynomial hierarchy, $\ph$, is defined by 
$\ph=\bigcup\limits_{k\geq 0} 
\sigmak$.
\end{enumerate}
\end{enumerate}
\end{definition}

So, for instance, $\Sigma_1^{\littlep}=\np$, $\sigmatwo=\np^{\np}$, and 
$\sigmathree=\np^{(\np^{\np})}$.
The boolean or difference hierarchy
is a concept used to study the 
structure within the boolean closure of a class $\calc$  
(the closure of $\calc$ with respect to the boolean operations 
$\wedge$, $\vee$, and negation).
It has particularly often 
been studied in terms of boolean
hierarchies built on the classes $\sigmak$, $k\geq 1$.

\begin{definition}
\begin{enumerate}
\item For sets of languages $\calcone$ and $\calctwo$, let
$\calcone \ominus \calctwo=\{L_1-L_2\condition L_1 \in \calcone \wedge 
L_2 \in \calctwo\}$.
\item
\cite{cai-gun-har-hem-sew-wag-wec:j:bh1,cai-gun-har-hem-sew-wag-wec:j:bh2} 
For all $k\geq 1$,\nopagebreak%
\begin{enumerate}%
\item $\diff_1(\sigmak)=\sigmak$.
\item For all $m\geq1$, $\diff_{m+1}(\sigmak)=\sigmak\ominus\diff_m(\sigmak)$.
\item The boolean or difference hierarchy over $\sigmak$, $\bh(\sigmak)$, 
is defined as $\bh(\sigmak)=\bigcup\limits_{m\geq 1} \diff_m(\sigmak)$. 
\end{enumerate}
\end{enumerate}
\end{definition}

For instance, $\diff_2(\np)$ is exactly the class 
${\rm DP}$~\cite{pap-yan:j:dp}. 
Similarly (and for reasons to 
be explained more formally later), 
we will for every $k\geq 1$ denote $\diff_2(\sigmak)$ by 
${\rm D}\!\cdot\!\sigmak$.
It is a well-known fact that,
for every $k\geq 1$, it holds that
$\bh(\sigmak)$, 
the difference hierarchy over $\sigmak$, 
is sandwiched 
between $\sigmak \cup \pik$ and $\Delta_{k+1}^{\littlep}$ 

Restricting the type of access to an oracle one has leads 
to the notions of bounded-Turing and bounded-truth-table query 
classes (see, e.g., Ladner, Lynch, and 
Selman~\cite{lad-lyn-sel:j:com}).

\begin{definition} Let $k\geq 0$ be an integer.
\begin{enumerate}
\item 
For $m\geq 1$, 
$\p^{\sigmak[m]}$ 
denotes the set of languages recognizable by some deterministic 
polynomial-time Turing machine (DPTM) making at most $m$ queries to a 
$\sigmak$ oracle.
\item
For $m\geq 1$, 
$\p^{\sigmak}_{m\hbox{-}{\rm tt}}$ 
denotes the set of languages recognizable by some DPTM 
making {\em in parallel\/} 
(at once, without knowing the answer of any query) 
at most $m$ queries to a $\sigmak$ oracle.
\item 
The bounded-query hierarchy and the bounded-truth-table hierarchy over 
$\sigmak$ are formed by the classes $\p^{\sigmak[m]}$ and 
$\p^{\sigmak}_{m\hbox{-}{\rm tt}}$, $m\geq 1$, respectively.
\end{enumerate}
\end{definition}
Obviously for all $k\geq 1$, 
$\p^{\sigmak}_{m\hbox{-}{\rm tt}}\subseteq \p^{\sigmak[m]}$ for all 
$m\geq 1$.  Also, $\p^{\sigmak}_{1\hbox{-}{\rm tt}}=\p^{\sigmak[1]}$.

It is well-known~\cite{koe-sch-wag:j:diff}
that the bounded-truth-table and 
the boolean hierarchy intertwine, i.e., 
for all $m\geq 1$ and all $k \geq 1$, 
$$
\diffmsigk \cup \co\diffmsigk \subseteq \p^{\sigmak}_{m\hbox{-}{\rm tt}}
\subseteq \diff_{m+1}(\sigmak) \cap \co\diff_{m+1}(\sigmak).
$$
Hence a result $\diffmplusonesigk=\codiffmplusonesigk \implies \ph=\calc$ 
yields as an easy corollary 
$\psigkmtt=\psigkmponett \implies \ph=\calc$.
We are now prepared to turn to the results obtained in the 
papers under consideration.

\paragraph{1) Kadin 1987~\cite{kad:toutdate:bh,kad:joutdatedbychangkadin:bh}}

Kadin started a line of research that studies the question of to what 
level the polynomial hierarchy collapses if the boolean hierarchy collapses. 
He showed that a collapse of the boolean hierarchy over $\sigmak$ at level $m$ 
implies a collapse of the polynomial hierarchy to its $(k+2)$nd level, 
$\sigmakptwo$.

\begin{theorem}~\cite{kad:toutdate:bh,kad:joutdatedbychangkadin:bh}
\label{t:k}
For all $m\geq 1$ and all $k\geq 1$, 
if ${\rm DIFF}_m(\sigmak)=\co{\rm DIFF}_m(\sigmak)$ then 
$\ph=\Sigma_{k+2}^{\littlep}$.
\end{theorem}

\begin{corollary}~\cite{kad:toutdate:bh,kad:joutdatedbychangkadin:bh}
For all $m\geq 1$ and all $k\geq 1$, 
if $\p^{\sigmak}_{m\hbox{-}{\rm tt}}=\p^{\sigmak}_{(m+1)\hbox{-}{\rm tt}}$ 
then $\ph=\Sigma_{k+2}^{\littlep}$.
\end{corollary}

\paragraph{2) Wagner 1987~\cite{wag:t:n-o-q-87version}}

Kadin's technique together with oracle replacement enabled 
Wagner to improve Kadin's results significantly by showing that a collapse 
of the boolean hierarchy over $\sigmak$ at level $m$ implies a collapse of the 
polynomial hierarchy to $\Delta_{k+2}^{\littlep}$.

\begin{theorem}~\cite{wag:t:n-o-q-87version}
\label{t:w1}
For all $m\geq 1$ and all $k\geq 1$, if $\diffmsigk=\co\diffmsigk$ then 
$\ph=\Delta_{k+2}^{\littlep}$.
\end{theorem}

\begin{corollary}~\cite{wag:t:n-o-q-87version}
For all $m\geq 1$ and all $k\geq 1$, 
if $\psigkmtt=\psigkmponett$ then 
$\ph=\Delta_{k+2}^{\littlep}$.
\end{corollary}

\paragraph{3) Wagner 1989~\cite{wag:t:n-o-q-89version}}

Wagner observed that a modified definition of hard strings yields an even 
stronger collapse of the polynomial hierarchy.
In particular, he showed that a  collapse 
of the boolean hierarchy over $\sigmak$ at level $m$ implies 
a collapse of the polynomial 
hierarchy to a level within $\Delta_{k+2}^{\littlep}$, 
namely, the boolean closure of $\sigmakpone$, $\bh(\sigmakpone)$.

\begin{theorem}~\cite{wag:t:n-o-q-89version}
\label{t:w2}
For all $m\geq 1$ and all $k\geq 1$, if $\diffmsigk=\co\diffmsigk$ then 
$\ph=\bh(\sigmakpone)$.
\end{theorem}

\begin{corollary}~\cite{wag:t:n-o-q-89version}
For all $m\geq 1$ and all $k\geq 1$, 
if $\psigkmtt=\psigkmponett$ then 
$\ph=\bh(\sigmakpone)$.
\end{corollary}

\paragraph{4) Chang/Kadin 1989~\cite{cha-kad:toutdated:bh,cha-kad:j:closer}}

Chang and Kadin refined the method originally used by 
Kadin to further tighten the connection between the boolean hierarchy and the 
polynomial hierarchy. 
Unaware of Wagner's work they improved his results. 
They showed that a  collapse 
of the boolean hierarchy over $\sigmak$ at level $m$ implies a 
collapse of the polynomial 
hierarchy to a level within the boolean closure 
of $\sigmakpone$, namely, the $m$th level of the boolean hierarchy 
over $\sigmakpone$.

\begin{theorem}~\cite{cha-kad:toutdated:bh,cha-kad:j:closer}
\label{t:ck}
For all $m\geq 1$ and all $k\geq 1$, if $\diffmsigk=\co\diffmsigk$ then 
$\ph=\diff_m(\sigmakpone)$.
\end{theorem}

\begin{corollary}~\cite{cha-kad:toutdated:bh,cha-kad:j:closer}
For all $m\geq 1$ and all $k\geq 1$, 
if $\psigkmtt=\psigkmponett$ then 
$\ph=\diff_{m+1}(\sigmakpone)$.
\end{corollary}

\paragraph{5) Beigel/Chang/Ogiwara 1991~\cite{bei-cha-ogi:tOUT:boolean,bei-cha-ogi:j:difference-hierarchies}}

Beigel, Chang, and Ogihara, 
while picking up ideas developed by Wagner, were 
able to draw a stronger conclusion.
In particular, they showed that a  collapse 
of the boolean hierarchy over $\sigmak$ at level $m$ implies a 
collapse of the polynomial 
hierarchy to a level within the $m$th level of the boolean hierarchy 
over $\sigmakpone$, 
namely, to $\left(\p^{\np}_{(m-1)\hbox{-}{\rm tt}}\right)^{\sigmak}$, 
the class of languages that can be accepted by some deterministic 
polynomial-time machine making at most $m-1$ parallel queries to an 
$\np^{\sigmak}=\sigmakpone$ oracle and an unlimited number of queries to a 
$\sigmak$ oracle.

\begin{theorem}~\cite{bei-cha-ogi:tOUT:boolean,bei-cha-ogi:j:difference-hierarchies}
\label{t:bco}
For all $m\geq 1$ and all $k\geq 1$, if $\diffmsigk=\co\diffmsigk$ then 
$\ph=\left({\p^{\np}_{(m-1)\hbox{-}{\rm tt}}}\right)^{\sigmak}$.
\end{theorem}

\begin{corollary}~\cite{bei-cha-ogi:tOUT:boolean,bei-cha-ogi:j:difference-hierarchies}
\label{c:bco}
For all $m\geq 1$ and all $k\geq 1$, 
if $\psigkmtt=\psigkmponett$ then 
$\ph=\left({\p^{\np}_{m\hbox{-}{\rm tt}}}\right)^{\sigmak}$.
\end{corollary}

\paragraph{6) Hemaspaandra/Hemaspaandra/Hempel 1996~\cite{hem-hem-hem:tOutBydown-sep:upward-sep}}

Motivated by the question of 
whether the collapse of query order classes 
has some effect on the polynomial hierarchy, Hemaspaandra, Hemaspaandra, and 
Hempel came up with a very surprising downward collapse result.
A collapse of the bounded-query hierarchy over $\sigmak$, $k>2$, 
at its first level implies a collapse of the polynomial hierarchy to $\sigmak$ 
itself; informally, 
the polynomial hierarchy collapses to a level that is below 
the level of the 
bounded-query hierarchy at which the initial collapse occurred.
This was the the first ``downward translation of equality'' 
(equivalently, ``downward collapse) result ever obtained within the 
bounded query hierarchies.

\begin{theorem}~\cite{hem-hem-hem:tOutBydown-sep:upward-sep}
\label{t:hhh1}
For $k>2$, if $\p^{\sigmak[1]}=\p^{\sigmak[2]}$ then $\ph=\sigmak=\pik$.
\end{theorem}

\paragraph{7) Hemaspaandra/Hemaspaandra/Hempel 1996~\cite{hem-hem-hem:tOutByConf:downward-translation,hem-hem-hem:jtoappear:downward-translation}}

Generalizing the ideas developed 
in~\cite{hem-hem-hem:tOutBydown-sep:upward-sep}, the authors extended 
their results to also hold for $j$-vs-$j+1$ queries. 
More precisely, a collapse of the bounded-truth-table hierarchy over 
$\sigmak$ at level $m$ 
implies the collapse of the boolean hierarchy over $\sigmak$ 
at level $m$.
This again is a downward collapse result as clearly 
$\diffmsigk\cup\co\diffmsigk \subseteq \psigkmtt$, 
and moreover the inclusion is believed to be strict.

\begin{theorem}~\cite{hem-hem-hem:tOutByConf:downward-translation,hem-hem-hem:jtoappear:downward-translation}
\label{t:hhh2}
For all $m\geq 1$ and all $k>2$, if $\psigkmtt=\psigkmponett$ then 
$\diffmsigk=\co\diffmsigk$.
\end{theorem}

This, together with the upcoming Theorem~\ref{t:new}, 
yields also a strong collapse of the polynomial hierarchy.

\begin{corollary} 
For all $m\geq 1$ and all $k>2$, if $\psigkmtt=\psigkmponett$ then
$\ph = \diff_m(\sigmak)\bolddelta\diff_{m-1}(\sigmakpone)$.
\end{corollary}

Note that the collapse of the polynomial hierarchy occurs, 
roughly speaking, one level lower in the boolean hierarchy over 
$\sigmakpone$ than could be concluded from the same hypothesis without 
Theorem~\ref{t:hhh2}. 

\paragraph{8) Buhrman/Fortnow 1996~\cite{buh-for:t:two-queries,buh-for:ctoappear:two-queries}}

Buhrman and Fortnow extended Theorem~\ref{t:hhh1} to the $k=2$ case;
they proved that $\p^{\sigmatwo[1]}=\p^{\sigmatwo[2]}$ implies 
a collapse of the polynomial hierarchy to $\sigmatwo$, 
establishing a downward collapse in the second level of the 
polynomial hierarchy.

\begin{theorem}~\cite{buh-for:t:two-queries}
\label{t:bf}
If $\p^{\sigmatwo[1]}=\p^{\sigmatwo[2]}$ then $\ph=\sigmatwo=\pitwo$.
\end{theorem}

\paragraph{9) Hemaspaandra/Hemaspaandra/Hempel 1997~\cite{hem-hem-hem:t:translating-downwards}}

In~\cite{hem-hem-hem:t:translating-downwards} the approaches 
of~\cite{hem-hem-hem:tOutByConf:downward-translation,hem-hem-hem:jtoappear:downward-translation}
and~\cite{buh-for:t:two-queries} were combined with new ideas to obtain a 
result that implies Theorems~\ref{t:hhh1},~\ref{t:hhh2},~\ref{t:bf}, 
and more. 

\begin{theorem}~\cite{hem-hem-hem:t:translating-downwards}
\label{t:hhh3}
For all $m\geq 1$ and all $k > 1$, 
if $\psigkmtt=\psigkmponett$ then $\diffmsigk=\co\diffmsigk$.
\end{theorem}

This is a very general downward collapse result, as the $m$th 
level of the boolean hierarchy over $\sigmak$ is contained in $\psigkmtt$.
In light of Theorem~\ref{t:new} the above Theorem~\ref{t:hhh3} also gives 
a collapse of the polynomial hierarchy that was previously unknown to hold.

\begin{corollary}
For all $m\geq 1$ and all $k > 1$, 
if $\psigkmtt=\psigkmponett$ then 
$\ph = \diff_m(\sigmak)\bolddelta\diff_{m-1}(\sigmakpone)$.
\end{corollary}

\section{Improving the Collapse of the Polynomial Hierarchy Under the 
Hypothesis that the Boolean Hierarchy Over \boldmath$\sigmak$\unboldmath\ 
Collapses}
\label{s:improved}

The first five papers of this survey obtained deeper and deeper collapses
of the polynomial hierarchy if the boolean hierarchy over $\sigmak$ collapses. 
The strongest result previously known is due to Beigel, Chang, and 
Ogihara~\cite{bei-cha-ogi:tOUT:boolean,bei-cha-ogi:j:difference-hierarchies}, 
see Theorem~\ref{t:bco}. Theorem~\ref{t:bco} says that, 
given a collapse of the boolean hierarchy over $\sigmak$ at level $m$, 
the polynomial hierarchy collapses to 
$\left({\p^{\np}_{(m-1)\hbox{-}{\rm tt}}}\right)^{\sigmak}$, 
a class contained in $\diff_m(\sigmakpone)$.

Define for complexity classes $\calc$ and $\cald$, 
$\calc\bolddelta\cald=\{C\Delta D \condition C\in \calc\wedge D \in \cald\}$, 
where $C\Delta D$ denotes the symmetric difference of the sets $C$ and $D$.
A careful analysis of the proof of Theorem~\ref{t:bco} as given
in~\cite{bei-cha-ogi:j:difference-hierarchies} in combination with a new 
trick, namely applying an idea developed 
in~\cite{bei-cha-ogi:tOUT:boolean,bei-cha-ogi:j:difference-hierarchies}  
twice, yields the following theorem. 
Theorem~\ref{t:new} has been independently obtained  
by Reith and Wagner~\cite{rei-wag:unpub:boolean-lowness}.

\begin{theorem}\label{t:new}
For all $m\geq 1$ and all $k\geq 1$, if $\diffmsigk=\co\diffmsigk$ then 
$\ph=\diffmsigk\bolddelta\diff_{m-1}(\sigmakpone)$.
\end{theorem}

Let us compare the results of Theorem~\ref{t:bco} and Theorem~\ref{t:new}.
Though both theorems 
collapse 
the polynomial hierarchy to a class containing $\diff_{m-1}(\sigmakpone)$ and 
contained in $\diff_{m}(\sigmakpone)$, 
their results differ substantially.
It is immediate from a recent paper of 
Wagner~\cite{wag:jtoappear:parallel-difference} that 
$\left({\p^{\np}_{(m-1)\hbox{-}{\rm tt}}}\right)^{\sigmak}$ 
is a strict superset of $\diffmsigk\bolddelta\diff_{m-1}(\sigmakpone)$ 
unless the polynomial hierarchy collapses.
Furthermore, observe that  
$\left({\p^{\np}_{(m-1)\hbox{-}{\rm tt}}}\right)^{\sigmak}$ 
involves $m-1$ parallel queries to a $\sigmakpone$ oracle and an 
unlimited number of queries to a $\sigmak$ oracle. 
So the $\p$ base machine of 
$\left({\p^{\np}_{(m-1)\hbox{-}{\rm tt}}}\right)^{\sigmak}$ 
evaluates $m-1$ bits of information originating from 
the parallel $\sigmakpone$ queries and polynomially many bits of information 
from the $\sigmak$ queries.
In contrast, $\diffmsigk\bolddelta\diff_{m-1}(\sigmakpone)$ involves 
just two bits of information, which are evaluated via a fixed truth-table, 
namely the XOR-truth-table. 
One bit of information comes from the $\diff_{m-1}(\sigmakpone)$ 
part consisting of  $m-1$ underlying parallel queries to a $\sigmakpone$ 
oracle evaluated with one fixed truth-table. 
The second bit of information, the one from the $\diffmsigk$ part, implicitly 
contains $m$ parallel queries to a $\sigmak$ oracle which again are 
evaluated via a fixed truth-table.
In a nutshell, we have improved from 
unlimited many queries to $\sigmak$ and 
$(m-1)$-truth-table queries to $\sigmakpone$,
to $m$-fixed-truth-table queries to $\sigmak$ and
$(m-1)$-fixed-truth-table queries to $\sigmakpone$. 

Before we are prove Theorem~\ref{t:new}, let us agree on the following 
convention: Whenever we talk about polynomials let us assume that those 
polynomials are of the form $n^a+b$ for some integers $a,b>0$. 
Since all complexity classes under consideration 
are closed under many-one reductions and 
the polynomials involved in the upcoming proofs 
(in this section as well as in Section~\ref{s:evolution}) 
always play the role of a function bounding the 
running time of some Turing machine or the 
length of some variable, we can make 
this assumption without loss of generality. 
This convention has the advantage that a polynomial $p$ now satisfies 
$p(n+1)>p(n)>n$ for all $n$, a condition we will need throughout our proofs.
Readers interested in the general flavor of the proof of 
Theorem~\ref{t:new} are encouraged 
to read the proof assuming $k=3$ and $m=2$.

{\bf Proof of Theorem~\ref{t:new}:}
The proof is structured in a way that the reader will easily find analogies 
to the proofs of the special cases in Subsection~\ref{ss:detailed}.

Observe that the claim is immediate for $m=1$. So suppose $m\geq 2$.
\begin{description}

\item[A] 
Let $\Sigma=\{0,1\}$ and let $\# \not\in \Sigma$.
Let $\langle . \rangle$ be a pairing function that maps sequences of 
length at most $m+1$ of strings over $\sigmastar\cup\{\#\}$ to $\sigmastar$ 
having the standard properties such as polynomial-time computability and 
invertibility etc.
Let $s$ be a polynomial bounding the size of $\langle . \rangle$, 
in particular let  
$|\pair{x_1,x_2,\dots,x_j}| \leq s(\max\{|x_1|,|x_2|,\dots ,|x_j|\})$ 
for all $1\leq j\leq m+1$ and all $x_1,x_2,\dots,x_j \in \sigmastar$. 
Define $s^{(0)}(n)=n$ and 
$s^{(j)}(n)=\underbrace{s(s(\cdots s}_{j~times}(n) \cdots ))$ for all $n$ 
and all $j\geq 1$.

For every $k\geq 1$, let $\lsigk$ be a many-one complete language for 
$\sigmak$ and hence 
$\lpik=\overline{\lsigk}$ is a complete language for $\pik$. 
Define $L_{\diff_1(\pik)}=\lpik$, and for every $m\geq 2$, 
$\ldiffmpik=\{\pair{x,y}\condition x\in \lpik \wedge 
y \not\in L_{\diff_{m-1}(\pik)}\}$. 
It is not hard to verify that for all $m\geq 1$, 
$\ldiffmpik$ is many-one complete for $\diffmpik$. 
So, $\overline{\ldiffmpik}$ is complete for $\co\diffmpik$ for all $m$.

Note also that $\diffmpik=\diffmsigk$ if $m$ is even and 
$\diffmpik=\co\diffmsigk$ if $m$ is odd. 
So in general, $$\diffmsigk=\co\diffmsigk \iff \diffmpik=\co\diffmpik.$$

\item[B] Suppose $\diffmsigk=\co\diffmsigk$. 
Hence $\diffmpik=\co\diffmpik$. Thus there exists a many-one reduction $h$ 
from $\ldiffmpik$ to $\overline{\ldiffmpik}$. 
In other words, for all $x_1,x_2 \in \sigmastar$,
$$\pair{x_1,x_2} \in \ldiffmpik \iff h(\pair{x_1,x_2})\not\in \ldiffmpik.$$

Let $h'$ and $h''$ be the polynomial-time computable functions 
such that for all $x_1,x_2 \in \sigmastar$,
$h(\pair{x_1,x_2})=\pair{h'(\pair{x_1,x_2}),h''(\pair{x_1,x_2})}$.
Hence, we have for all $x_1,x_2 \in \sigmastar$, 
$$(*)\qquad x_1\in \lpik\wedge x_2 \notin L_{\diff_{m-1}(\pik)} \iff 
h'(\pair{x_1,x_2}) \notin \lpik \vee h''(\pair{x_1,x_2}) \in 
L_{\diff_{m-1}(\pik)}.$$

\item[C] Recall that we want to show a collapse of the polynomial hierarchy. 
Though we do not claim that we can prove $\sigmak=\pik$ we will nevertheless 
show that a $\sigmak$ algorithm for $\lpik$ exists which
requires certain additional input.
We will extend this to also give $\sigmak$ algorithms for $\lsigkpone$ and 
$\lsigkptwo$, both algorithms requiring additional input.

Let $n$ be an integer.
In light of the equivalence (*) we call 
the string $x_1$ $m$-easy for length $n$ 
if and only if $|x_1|\leq n$ and 
$(\exists x_2~|x_2|\leq s^{(m-2)}(n))[h'(\pair{x_1,x_2}) \not\in \lpik]$. 
Clearly, if $x_1$ is $m$-easy for length $n$ then $x_1 \in \lpik$.

A string $x_1$ is said to be $m$-hard for length $n$ 
if and only if $|x_1|\leq n$, $x_1 \in \lpik$, and 
$(\forall x_2~|x_2|\leq s^{(m-2)}(n))[h'(\pair{x_1,x_2}) \in \lpik]$. 
It is not hard to verify that the strings in 
${(\lpik)}^{\leq n}$ divide into $m$-easy and $m$-hard 
strings for length $n$. 

\begin{description}
\item[Case 1] 
There are no $m$-hard strings for length $n$.\\
Hence all strings in ${(\lpik)}^{\leq n}$ are $m$-easy for length $n$. 
Thus deciding whether $x$, $|x|=n$, is in $\lpik$ is equivalent 
to deciding whether $x$ is $m$-easy for length $n$. 
Note that the latter can be done by the following $\sigmak$ algorithm:
\begin{enumerate}
\item Guess $y$, $|y| \leq s^{(m-2)}(n)$.
\item Compute $h(\pair{x,y})$.
\item Accept if and only if $h'(\pair{x,y}) \not\in \lpik$.
\end{enumerate}
\item[Case 2] 
There exist $m$-hard strings for length $n$.\\
Let $\omega_m$ be an $m$-hard string for length $n$, hence $|\omega_m|\leq n$. 
Let $h_{(\omega_m)}$ be the function such that for  every $u \in \sigmastar$, 
$h_{(\omega_m)}(u)=h''(\pair{\omega_m,u})$. 
Note that given $\omega_m$, $h_{(\omega_m)}(u)$ is computable in 
time polynomial in $\max\{n,|u|\}$. 
According to the definition of $m$-hard strings and equivalence (*) we have
for all $u$, $|u|\leq s^{(m-2)}(n)$, 
$$
u\in L_{\diff_{m-1}(\pik)} \iff  h_{(\omega_m)}(u) 
\not\in L_{\diff_{m-1}(\pik)}.
$$
Thus we have a situation similar to the one in {\bf B} but $m$ replaced by 
$m-1$ and also the equivalence holds only for an initial segment.
Let $h'_{(\omega_m)}$ and $h''_{(\omega_m)}$ be the functions such that 
for all $x_1,x_2 \in \sigmastar$,
$h_{(\omega_m)}(\pair{x_1,x_2})=
\pair{h'_{(\omega_m)}(\pair{x_1,x_2}),h''_{(\omega_m)}(\pair{x_1,x_2})}$.

Let $u=\pair{u_1,u_2}$.
Hence for all
$u_1$, $|u_1|\leq n$, and all $u_2$, $|u_2|\leq s^{(m-3)}(n)$,
\begin{eqnarray*}
\lefteqn{u_1 \in \lpik \wedge u_2 \not\in L_{\diff_{m-2}(\pik)} \iff 
}\hspace{4cm}\\
&& h'_{(\omega_m)}(\pair{u_1,u_2})  \not\in \lpik \vee 
h''_{(\omega_m)}(\pair{u_1,u_2}) \in L_{\diff_{m-1}(\pik)}.
\end{eqnarray*}

We call the string $u_1$ $(m-1)$-easy for length $n$ 
if and only if $|u_1|\leq n$ and 
$(\exists u_2~|u_2|\leq s^{(m-3)}(n))
[h'_{(\omega_m)}(\pair{u_1,u_2}) \not\in \lpik]$. 
If $u_1$ is $(m-1)$-easy for length $n$ then $u_1 \in \lpik$.

A string $u_1$ is said to be $(m-1)$-hard for length $n$ 
if and only if $|u_1|\leq n$, $u_1 \in \lpik$, and 
$(\forall u_2~|u_2|\leq s^{(m-3)}(n))
[h'_{(\omega_m)}(\pair{u_1,u_2}) \in \lpik]$. 

It is not hard to verify that, given an $m$-hard string $\omega_m$ 
for length $n$, the strings 
in ${(\lpik)}^{\leq n}$ divide into $(m-1)$-easy and $(m-1)$-hard 
strings for length $n$. 
Note that $(m-1)$-hardness is only defined with respect to some particular 
$m$-hard string $\omega_m$.

\begin{description}
\item[Case 2.1] 
There exist no $(m-1)$-hard strings for length $n$.\\
Hence similar to Case 1, 
all strings in ${(\lpik)}^{\leq n}$ are $(m-1)$-easy for length $n$,  
deciding whether $x$, $|x|=n$, is in $\lpik$ is equivalent 
to deciding whether $x$ is $(m-1)$-easy for length $n$
which, with the help of $\omega_m$,  can be done with a $\sigmak$ algorithm.
\item[Case 2.2] 
There exist $(m-1)$-hard strings for length $n$.\\
Let $\omega_{m-1}$ be an $(m-1)$-hard string for length $n$, 
$|\omega_{m-1}|\leq n$. 
Let $h_{(\omega_m,\omega_{m-1})}$ be the function such that for all 
$v \in \sigmastar$, 
$h_{(\omega_m,\omega_{m-1})}(v)=h''_{(\omega_m)}(\pair{\omega_{m-1},v})$. 
Note that given $\omega_m$ and $\omega_{m-1}$, 
$h_{(\omega_m,\omega_{m-1})}(v)$ 
is computable in time polynomial in $\max\{n,|v|\}$.
Hence, for all $v$, $|v|\leq s^{(m-3)}(n)$, 
$$
v\in L_{\diff_{m-2}(\pik)} \iff 
h_{(\omega_m,\omega_{m-1})}(v) \not\in L_{\diff_{m-2}(\pik)}.
$$

Continuing in that manner we define for $\ell \geq 2$, $\ell$-hard and 
$\ell$-easy strings for length $n$. 
Note that these terms are defined with respect to some fixed 
$m$-hard,$(m-1)$-hard, \dots, $(\ell+1)$-hard strings. 
In other words, a string is only $\ell$-hard or $\ell$-easy with respect to 
a particular sequence of hard strings 
$\omega_m,\omega_{m-1},\dots,\omega_{\ell+1}$. 
We define that there are no 1-hard strings for length $n$, 
and a string $z$ is called 1-easy for length $n$ if and only if $|z|\leq n$ 
and $h_{(\omega_m,\omega_{m-1},\dots,\omega_2)}(z) \not\in \lpik$.
\end{description}
\end{description}

A sequence $\omega_m,\omega_{m-1},\dots,\omega_{\ell}$, $\ell \geq 2$, 
is called a hard sequence for length $n$ if and only if 
$\omega_j$ is 
$j$-hard (with respect to $\omega_m$, $\omega_{m-1}$, \dots, $\omega_{j+1}$) 
for length $n$ for all $j$, $\ell\leq j \leq m$.
We call $m-\ell+1$ the order of the hard sequence 
$\omega_m,\omega_{m-1},\dots,\omega_{\ell}$. 

A sequence $\omega_m,\omega_{m-1},\dots,\omega_{\ell}$ is called 
a maximal hard sequence for length $n$ if and only if 
$\omega_m,\omega_{m-1},\dots,\omega_{\ell}$ is a hard sequence for 
length $n$ and 
there are no $(\ell -1)$-hard strings (with respect to 
$\omega_m$, $\omega_{m-1}$, \dots, $\omega_{\ell}$) for length $n$.
As a special case, 
$\#$ is called a maximal hard sequence for length $n$ if and only if 
there exists no $m$-hard string for length $n$, 
$\#$ is said to have order zero.
Note that deciding whether, given a sequence of strings $s$ and 
an integer $n$, $s$ is a hard sequence for length $n$ can be done with a 
$\pik$ algorithm.

It is clear that for every $n$, a maximal hard sequence for length $n$ 
always exists and has order at most $m-1$ since there are no 1-hard strings 
for length $n$.

\item[D] One maximal hard sequence is needed to reduce part of $\lpik$ 
to a $\sigmak$ language.

\item[\phantom{D}]
{\em Claim~D: There exists a set $A \in \sigmak$ such that 
for all $x\in \sigmastar$ and all $l\geq |x|$, 
if $\omega_m,\omega_{m-1},\dots,\omega_{\ell}$ is a maximal hard 
sequence for length $l$ then 
$$x \in \lpik \iff 
\pair{x,1^l,\omega_m,\omega_{m-1},\dots,\omega_{\ell}} \in A.$$
}

Let $x\in \sigmastar$ and let 
$\omega_m,\omega_{m-1},\dots,\omega_{\ell}$ be a maximal hard 
sequence for length $l$, $l\geq |x|$. 
Note that $\ell \geq 2$. 
Since $\omega_m,\omega_{m-1},\dots,\omega_{\ell}$ is maximal hard, 
no string of length less or equal to $l$ is $(\ell -1)$-hard with respect to 
$\omega_m,\omega_{m-1},\dots,\omega_{\ell}$. 
Hence, for every string $y$, $|y|\leq l$,
$y \in (\lpik)^{\leq l}$ if and only if $y$ is $(\ell -1)$-easy for length $l$.
This holds especially for $x$ itself (recall $|x|\leq l$).
But testing whether $x$ is $(\ell -1)$-easy for length $l$ 
can clearly be done by a $\sigmak$ algorithm when receiving $x$, $1^l$, and  
$\omega_m,\omega_{m-1},\dots,\omega_{\ell}$ as inputs.
In particular, define 
$A=\{\pair{x,1^l,\omega_m,\omega_{m-1},\dots,\omega_{\ell}}\condition 
(\ell=2 \wedge h_{(\omega_m,\omega_{m-1},\dots,\omega_{\ell})}(x) 
\not\in \lpik) \vee 
(\ell >2 \wedge (\exists y~|y|\leq s^{\ell -3}(l))
[h'_{(\omega_m,\omega_{m-1},\dots,\omega_{\ell})}(\pair{x,y}) 
\not\in \lpik]\}$.

\item[E]
One maximal hard sequence for sufficiently large length also suffices to 
give a reduction from some of $\lsigkpone$ to a $\sigmak$ language.

\item[\phantom{E}]
{\em Claim~E: There exist a set $B\in \sigmak$ and a polynomial $q$ 
such that for all $x\in \sigmastar$ and all $l\geq q(|x|)$, 
if $\omega_m,\omega_{m-1},\dots,\omega_{\ell}$ is a maximal hard 
sequence for length $l$ then 
$$x \in \lsigkpone \iff 
\pair{x,1^l,\omega_m,\omega_{m-1},\dots,\omega_{\ell}} \in B.$$
}

Let $p$ be polynomial such that for all $x\in \sigmastar$, 
$$
x\in \lsigkpone \iff (\exists y~|y|\leq p(|x|))[\pair{x,y} \in \lpik].
$$
Applying Claim~D we obtain that there is a set $A\in \sigmak$ such that 
for all $x$ and all $l\geq s(p(|x|))$, 
if $\omega_m,\omega_{m-1},\dots,\omega_{\ell}$ is a 
maximal hard sequence for length $l$ then
$$
x\in \lsigkpone \iff (\exists y~|y|\leq p(|x|))
[\pair{\pair{x,y},1^l,\omega_m,\omega_{m-1},\dots,\omega_{\ell}} \in A].
$$
Note that the right-hand-side of the above equivalence clearly defines a 
$\sigmak$ language $B$. 
Define $q$ to be a polynomial such that $q(n) \geq s(p(n))$ for all $n$.
This proves the claim.

\item[F]
In contrast to {\bf D} and {\bf E}, two maximal hard sequences 
for different length are required when reducing some of $\lsigkptwo$ 
to a $\sigmak$ language.

\item[\phantom{F}]
{\em Claim~F: There exist a set $C\in \sigmak$ and polynomials $q_1$, 
$q_2$ 
such that for all $x\in \sigmastar$,  
if $\omega_m,\omega_{m-1},\dots,\omega_{\ell}$ and 
$\omega'_m,\omega'_{m-1},\dots,\omega'_{\ell'}$ are maximal hard 
sequences for length $q_1(|x|)$ and $q_2(|x|)$, respectively, then 
$$x \in \lsigkptwo \iff 
\pair{x,\pair{\omega_m,\omega_{m-1},\dots,\omega_{\ell}},
\pair{\omega'_m,\omega'_{m-1},\dots,\omega'_{\ell'}}} \in C.$$
}

Let $p'$ be a polynomial such that for all $x\in \sigmastar$,
$$
x\in \lsigkptwo \iff (\exists y~|y|\leq p'(|x|))
[\pair{x,y} \not\in \lsigkpone].
$$
Applying Claim~E we obtain that there is a set $B\in \sigmak$ and 
a polynomial $q$ such that for all $x \in \sigmastar$ and all 
$l\geq q(s(p'(|x|)))$, 
if $\omega_m,\omega_{m-1},\dots,\omega_{\ell}$ is a maximal 
hard sequence for length $l$ then 
$$
x\in \lsigkptwo \iff (\exists y~|y|\leq p'(|x|))
[\pair{\pair{x,y},1^l,\omega_m,\omega_{m-1},\dots,\omega_{\ell}} \not\in B].
$$
Define $q_1$ to be a polynomial such that $q_1(n)\geq q(s(p'(n)))$ 
for all $n$.
Define $L'=\{\pair{x,1^l,\omega_m,\omega_{m-1},\dots,\omega_{\ell}}
\condition (\exists y~|y|\leq p'(|x|))
[\pair{\pair{x,y},1^l,\omega_m,\omega_{m-1},\dots,\omega_{\ell}} 
\not\in B]\}$.
Note that $L' \in \sigmakpone$ and let $g$ be a many-one reduction from 
$L'$ to $\lsigkpone$. 
Hence we have for all $x \in \sigmastar$, 
if $\omega_m,\omega_{m-1},\dots,\omega_{\ell}$ is a maximal 
hard sequence for length $q_1(|x|)$ then
$$
x\in \lsigkptwo \iff 
g(\pair{x,1^{q_1(|x|)},\omega_m,\omega_{m-1},\dots,\omega_{\ell}}) 
\in \lsigkpone.
$$
Applying Claim~E for the second time we obtain that 
for all $x \in \sigmastar$,  
if $\omega_m,\omega_{m-1},\dots,\omega_{\ell}$ is a maximal 
hard sequence for length $q_1(|x|)$ and 
$\omega'_m,\omega'_{m-1},\dots,\omega'_{\ell'}$ is a maximal hard 
sequence for length $l$,
$l\geq q(|g(\pair{x,1^{q_1(|x|)},\omega_m,\omega_{m-1},\dots,\omega_{\ell}})|)$, 
then 
$$
x\in \lsigkptwo \iff 
\pair{g(\pair{x,1^{q_1(|x|)},\omega_m,\omega_{m-1},\dots,\omega_{\ell}}),1^l,
\omega'_m,\omega'_{m-1},\dots,\omega'_{\ell'}} \in B.
$$
Let $\wh{q}$ be a polynomial such that $|g(z)|$ is bounded by $\wh{q}(|z|)$ 
for all $z$. 
Define $q_2$ to be a polynomial such that 
$q_2(n) \geq q(\wh{q}(s(q_1(n))))$ for all $n$. 
Set 
\begin{eqnarray*}
\lefteqn{C=\{\pair{x,\pair{\omega_m,\omega_{m-1},\dots,\omega_{\ell}},
\pair{\omega'_m,\omega'_{m-1},\dots,\omega'_{\ell'}}}\condition}\hspace{4cm}\\ 
&& \pair{g(\pair{x,1^{q_1(|x|)},\omega_m,\omega_{m-1},\dots,\omega_{\ell}}),
1^{q_2(|x|)},
\omega'_m,\omega'_{m-1},\dots,\omega'_{\ell'}} \in B\}
\end{eqnarray*} 
and note that clearly $C \in \sigmak$.
This proves the claim.

\item[G] Applying the so called mind change technique in light of Claim~C 
yields that $\lsigkptwo \in \sigmak \bolddelta \diff_{2m-2}(\sigmakpone)$.

\item[\phantom{G}]
{\em Claim~G: $\ph \subseteq \sigmak \bolddelta 
\diff_{2m-2}(\sigmakpone)$.}

To prove the claim it suffices to give a 
$\sigmak \bolddelta \diff_{2m-2}(\sigmakpone)$ algorithm for $\lsigkptwo$.

A few definitions will be helpful. 
For sequences of strings $u=(u_1,u_2,\dots,u_j)$ and 
$v=(v_1,v_2,\dots,v_{j'})$,
$v$ is called an extension of $u$ if and only if $j\leq j'$ and for all 
$1\leq i \leq j$, $u_i=v_i$.  
$v$ is called a proper extension of $u$ if and only if 
$v$ is an extension of $u$ and $j < j'$.
A similar definition is made for pairs of sequences of strings.
For $(u,v)$ and $(u',v')$, where $u,u',v,v'$ are sequences of 
strings we call 
$(u',v')$ an extension of $(u,v)$ if and only if $u'$ is 
an extension of $u$ and $v'$ is an extension of $v$.
$(u',v')$ is called a proper extension of $(u,v)$ if and only if
$(u',v')$ is an extension of $(u,v)$, and $u'$ or $v'$ is a proper 
extension of $u$ or $v$, respectively. 

Let $\ell_1(n), \ell_2(n)$ to be the orders of the 
longest maximal hard sequences for lengths $q_1(n)$ and $q_2(n)$, 
respectively, where $q_1$ and $q_2$ are the polynomials spoken of in Claim~F. 
According to Claim~F, for all $x$,
$x\in \lsigkptwo$ if and only if there exist two hard sequences $s_1$ and
$s_2$ for length $q_1(|x|)$ and $q_2(|x|)$ of order $\ell_1(|x|)$ and 
$\ell_2(|x|)$, respectively, such that $\pair{x,s_1,s_2} \in C$.

Define 
$Q_0=\{x\condition \pair{x,\pair{\#},\pair{\#}} \in C\}$.
Define for $1\leq j$,

$Q_j=\{x\condition$
\begin{minipage}[t]{13cm}
there exist $r_1,s_1$, $r_2,s_2$,\dots,$r_j,s_j$ such that 
\begin{enumerate}
\item for all $1\leq i\leq j$, $r_i$ and $s_i$ are hard sequences for length 
$q_1(|x|)$ and $q_2(|x|)$, respectively,
\item for all $1\leq i\leq j-1$, $(r_{i+1},s_{i+1})$ is a proper 
extension of $(r_{i},s_{i})$, and 
\item 
$\chi_C(\pair{x,\pair{\#},\pair{\#}})\not= 
\chi_C(\pair{x,\pair{r_1},\pair{s_1}}$ 
and for all $1\leq i\leq j-1$, 
$\chi_C(\pair{x,\pair{r_i},\pair{s_i}})\not=
\chi_C(\pair{x,\pair{r_{i+1}},\pair{s_{i+1}}})\}$.
\end{enumerate}
\end{minipage}  

Observe that $Q_0 \in \sigmak$ and $Q_j \in \sigmakpone$, $1 \leq j$.
Since all hard sequences have order at most $m-1$ 
(and thus $\ell_1(n)\leq m-1$ and $\ell_2(n)\leq m-1$) and point 3 in the 
definition 
of $Q_j$ requires $(r_1,s_1)\not=(\#,\#)$ we obtain that 
for all $j > 2m-2$, $Q_j=\emptyset$.
Furthermore, it is not hard to verify that for all $x \in \sigmastar$,
$$
x\in \lsigkptwo \iff x \in Q_0 \Delta 
(Q_1-(Q_2-(\cdots -(Q_{2m-3}-Q_{2m-2})\cdots))).
$$
This shows
$\lsigkptwo \in \sigmak\bolddelta \diff_{2m-2}(\sigmakpone)$.

\item[H]
Applying the mind change technique to the result of Claim~G 
while exploiting Claim~E yields 
$\lsigkptwo \in \diff_{2m-1}(\sigmak) \bolddelta \diff_{m-1}(\sigmakpone)$.

\item[\phantom{H}]
{\em Claim~H: $\ph \subseteq \diff_{2m-1}(\sigmak) \bolddelta 
\diff_{m-1}(\sigmakpone)$.}

In light of Claim~G  and the fact that 
$\sigmak\bolddelta\diff_{2m-2}(\sigmakpone) 
\subseteq \diff_{2m-1}(\sigmakpone)$
it suffices to show 
$\diff_{2m-1}(\sigmakpone) \subseteq \diff_{2m-1}(\sigmak) \bolddelta 
\diff_{m-1}(\sigmakpone)$.

Let $L \in \diff_{2m-1}(\sigmakpone)$, hence there exist sets 
$L_1,L_2,\dots,L_{2m-1} \in \sigmakpone$ such that 
$L=L_1-(L_2-(\dots-(L_{2m-2}-L_{2m-1})\cdots))$.
According to Claim~E (note that Claim~E can be easily extended to hold for all 
$\sigmakpone$ languages and not just for $\lsigkpone$) there exist sets 
$B_1,B_2,\dots,B_{2m-1} \in \sigmak$ and polynomials $p_1,p_2,\dots,p_{2m-1}$ 
such that for all $x \in \sigmastar$ and all $1\leq i \leq 2m-1$, 
if $\omega^i_m,\omega^i_{m-1},\dots,\omega^i_{\ell_i}$ is a 
maximal hard sequence for length $l_i$,  $l_i\geq p_i(|x|)$, then 
$$x \in L_i \iff 
\pair{x,1^{l_i},\omega^i_m,\omega^i_{m-1},\dots,\omega^i_{\ell}} \in B_i.
$$ 
Let $\wh{p}$ be a polynomial such that $\wh{p}(n)\geq p_i(n)$ for all $n$ and 
all $1\leq i \leq 2m-1$. 
Define $D=\{\pair{x,\omega_m,\omega_{m-1},\dots,\omega_{\ell}}\condition 
\pair{x,1^{\wh{p}(|x|)},\omega_m,\omega_{m-1},\dots,\omega_{\ell}} \in 
B_1-(B_2-(\dots-(B_{2m-2}-B_{2m-1})\cdots))\}$.
Note that $D \in \diff_{2m-1}(\sigmak)$. 
We have for all $x \in \sigmastar$, 
if $\omega_m,\omega_{m-1},\dots,\omega_{\ell}$ is a maximal hard sequence 
for length $\wh{p}(|x|)$ then 
$$
x \in L \iff 
\pair{x,\omega_m,\omega_{m-1},\dots,\omega_{\ell}}\in D.
$$ 
Now we use a similar idea as in the proof of Claim~G. 
In particular, recall the definitions from the beginning of its proof. 
Define 
$P_0=\{x\condition \pair{x,\#} \in D\}$.
Define for $1\leq j$,

$P_j=\{x\condition$
\begin{minipage}[t]{13cm}
there exist $s_1$,$s_2$,\dots,$s_j$ such that 
\begin{enumerate}
\item for all $1\leq i\leq j$, $s_i$ is a  hard sequence for length 
$\wh{p}(|x|)$,
\item for all $1\leq i\leq j-1$, $s_{i+1}$ is a proper extension of 
$s_{i}$, and 
\item $\chi_{D}(\pair{x,\#})\not= 
\chi_{D}(\pair{x,s_1}$ and for all 
$1\leq i\leq j-1$, 
$\chi_{D}(\pair{x,s_i})\not=
\chi_{D}(\pair{x,s_{i+1}})\}$.
\end{enumerate}
\end{minipage}  

Note that $P_0 \in \diff_{2m-1}(\sigmak)$ and $P_j \in \sigmakpone$, $1\leq j$.
Since all hard sequences have order at most $m-1$ and point 3 in the 
definition of $P_j$ requires $s_1\not=\#$ we obtain that 
for all $j > m-1$, $P_j=\emptyset$.
Thus it is not hard to verify that for all $x \in \sigmastar$,
$$
x\in L \iff x \in P_0 \Delta 
(P_1-(P_2-(\cdots -(P_{m-2}-P_{m-1})\cdots))).
$$
Hence 
$L \in \diff_{2m-1}(\sigmak)\bolddelta \diff_{m-1}(\sigmakpone)$.
This completes the proof of Claim~H.

\item[I]
Applying the mind change technique again, this time to the result of Claim~H, 
gives the claim of the theorem being proven.

\item[\phantom{I}]
{\em Claim~I: $\ph \subseteq \diff_m(\sigmak) \bolddelta 
\diff_{m-1}(\sigmakpone)$.}

Observe that 
$\diff_{2m-1}(\sigmak) \bolddelta \diff_{m-1}(\sigmakpone) \subseteq 
\diff_m(\sigmakpone)$.  
In light of Claim~H it suffices to show
$\diff_{m}(\sigmakpone) \subseteq \diff_m(\sigmak) \bolddelta 
\diff_{m-1}(\sigmakpone)$ 
which can be done quite analogous to the proof of Claim~H.

\end{description}

\qed

\section{The Evolution of the Easy-Hard Technique}
\label{s:evolution}

This section is structured as follows.
Section~\ref{ss:nutshell} gives a short, non-technical, exceedingly 
informal summary of the technical contribution of each of the nine papers.
Section~\ref{ss:detailed} gives a detailed explanation of the easy-hard 
technique, and of the specific technical advances made by each of the 
nine papers.

\subsection{In a Nutshell}
\label{ss:nutshell}

The term easy-hard originates from Kadin's observation that in case 
that the 
boolean hierarchy collapses at level $m$ the strings of any particular 
length $n$ in a $\conp$ complete language divide into easy and hard strings. 
Hard strings are strings that allow one to 
translate a collapse of the boolean hierarchy from level $m$ 
to level $m-1$ in a restricted sense. 
Several hard strings eventually allow one to reduce a $\conp$ predicate to a 
$\np$ predicate.
In contrast, if no hard string at length $n$ exists then all strings 
of length $n$ in the $\conp$ complete language are easy 
and this allows one 
to directly reduce a $\conp$ predicate to a $\np$ predicate.
So, if we know whether there exist hard strings or not, 
and if, in case they exist, we are able to effectively compute them, 
we can with their help reduce a $\conp$ predicate to a $\np$ predicate and 
eventually collapse the polynomial hierarchy. 
This approach is central in each of the first five papers 
of our nine-paper survey.
The major difference among the five papers, 
and the main reasons for the difference in their results, 
is the way in which one obtains the needed 
information about the hard strings (their existence and the strings itself), 
and in which way one uses this information to collapse the 
polynomial hierarchy.
The last four papers of our survey use the (modified) easy-hard method to 
prove downward collapse results within the polynomial hierarchy.

Kadin~\cite{kad:joutdatedbychangkadin:bh} constructed a sparse set $S$ 
containing enough information such that 
one can effectively extract a lexicographically extreme hard string 
for a given length if there exists one.
He shows that $\conp \subseteq \np^S$, which by a result of 
Yap~\cite{yap:j:advice} implies $\ph\subseteq 
\sigmathree$.\footnote{In his original work 
Kadin claimed that his sparse set $S$ 
is even contained in $\sigmatwo$ (a claim he retracted later), 
which would have allowed him to conclude 
$\ph=\Theta_{k+2}^{\littlep}$.}

Wagner used a different approach in his two 
papers~\cite{wag:t:n-o-q-87version,wag:t:n-o-q-89version}. 
He collapsed the polynomial hierarchy directly 
(without constructing a sparse oracle) 
using oracle replacement and hard strings in 
the form of advice. 
The main reason for the stronger result in his second paper 
is a modified definition of easy and hard strings. 
Thus, instead of hard strings 
giving a reduction for only the strings of one particular length 
(and thus one needs a hard string for each length 
when collapsing the polynomial hierarchy),
Wagner's new definition yields that hard strings can give a reduction for all 
strings of length below a particular threshold.

Chang and Kadin~\cite{cha-kad:j:closer}, 
independent of Wagner's work, 
also used the stronger notion of hardness.
The observation that hard strings of larger length allow 
one to effectively gain 
information about the existence of hard strings at lower length together 
with  an elegant use of the nested difference structure of the 
boolean hierarchy over $\sigmatwo$, leads to their final result.

Beigel, Chang, and Ogihara~\cite{bei-cha-ogi:j:difference-hierarchies} 
further improved the results of Chang and Kadin.
They followed the approach of Wagner, but with two major  
innovations: 
First, they used complete languages for the levels of the boolean hierarchy 
that do not force them to distinguish between odd and even levels.
Second, they made use of the mind change technique to effectively 
check the existence 
of hard strings and their effect on the outcome of the reduction using 
those hard strings.
Their second innovation includes a modified argumentation 
for collapsing the polynomial hierarchy.

The key contribution of~\cite{hem-hem-hem:tOutBydown-sep:upward-sep}
was the insight that, since each string is easy or hard, one can 
completely discard 
the search for such strings!
Rather, one can simply always use the input itself as an easy or hard 
string (whichever it happened to be).
\cite{hem-hem-hem:tOutByConf:downward-translation} 
and~\cite{buh-for:t:two-queries} extended this approach in two different 
directions.
The approach of~\cite{hem-hem-hem:tOutBydown-sep:upward-sep} works only 
for 1-vs-2 query access to $\sigmak$, $k>2$.
\cite{buh-for:t:two-queries} extends the result to the 1-vs-2 query case for 
$k\geq 2$, via modifying the test of whether the input is easy or hard to 
nondeterministically simply assume both that the input is easy and that the 
input is hard. They add a new bit of ``code'' to ensure that the 
nondeterministic branch making the wrong assumption will do no harm to the 
overall algorithm.
\cite{hem-hem-hem:jtoappear:downward-translation}, on the other hand, removes 
the 1-vs-2 restriction but adds a scheme implementing 
``0-bit communication'' between machines.
They do this by having the machines independently latch onto a certain 
lexicographically extreme string signaled by the input.

However, recall from Section~\ref{s:results} the two improvements just 
mentioned---from $k>2$ to $k\geq 2$ via~\cite{buh-for:t:two-queries} and 
from 1-vs-2 to $j$-vs-$j+1$ 
via~\cite{hem-hem-hem:jtoappear:downward-translation}---are incomparable.
Neither paper allows both improvements to work simultaneously.
However, this was achieved in~\cite{hem-hem-hem:t:translating-downwards}, 
via a new twist.
\cite{hem-hem-hem:t:translating-downwards} provides an improved way of 
allowing the underlying $\sigmak$ machines of $\diff(\sigmak)$ languages 
to work together.
In particular, \cite{hem-hem-hem:t:translating-downwards} does so by 
exploiting the so-called telescoping normal form of boolean (or difference)  
hierarchies~\cite{cai-gun-har-hem-sew-wag-wec:j:bh1}---a normal form that 
in concept dates as far back as the work of Hausdorff~\cite{hau:b:sets}.

\subsection{Detailed Technical Discussion of the Easy-Hard Technique and its 
Extensions}
\label{ss:detailed}
 
In this section we study the easy-hard technique in its original version, 
and in the increasingly strong extensions developed in the series of papers 
that this article studies.
We will in the following prove for each of the papers 
just a special 
case of the key result obtained.
In particular we will start from one of the assumptions 
$\psigthreeone=\psigthreetwo$, $\psigthreetwott=\psigthreethreett$,
$\psigtwoone=\psigtwotwo$, or $\psigtwotwott=\psigtwothreett$ and 
prove the collapse of the polynomial hierarchy obtained by the corresponding 
paper. 

Let $\Sigma=\{0,1\}$ and let $\# \not\in \Sigma$ be a new symbol. 
Let $\pair{\cdots}$ be a pairing function mapping 
finite sequences of strings  from 
$\sigmastar\cup\{\#\}$ to $\sigmastar$,
and let this function have the 
standard properties, such as for instance being polynomially computable and 
invertible.
Let $s$ be a polynomial such that for all $x,y,z \in \sigmastar\cup\{\#\}$,
$|\pair{x,y}|\leq s(\max\{|x|,|y|\})$ and 
$|\pair{x,y,z}|\leq s(\max\{|x|,|y|,|z|\})$.
Recall our convention regarding polynomials from Section~\ref{s:improved}, 
in particular, 
since the polynomials involved in the upcoming proofs always play the role of 
a function bounding the 
running time of some Turing machine or the 
length of some variable, we without loss of generality assume that 
such polynomials always have the form $n^a+b$ for some integers $a,b>0$. 
This, for example, guarantees that for a polynomial $p$ it now holds that
$p(n+1)>p(n)>n$ for all $n$.

\paragraph{1) Kadin 1987~\cite{kad:toutdate:bh,kad:joutdatedbychangkadin:bh}}

\begin{theorem}\label{tt:kad}
If $\psigthreeone=\psigthreetwo$ then $\ph=\sigmafive$.
\end{theorem}

\proof
\begin{description}
\item[{\em Main Claim}] 
{\em If $\dsigthree=\codsigthree$ then $\ph=\sigmafive$,}

where $\dsigthree$ is the $\sigmathree$ analogue of the class 
DP~\cite{pap-yan:j:dp}, in particular $\dsigthree=\{L_1-L_2\condition 
L_1,L_2 \in \sigmathree\}$.

Since $\psigthreeone \subseteq \dsigthree \subseteq \psigthreetwo$ and 
$\psigthreeone \subseteq \codsigthree \subseteq \psigthreetwo$ 
the theorem follows immediately from the above claim.
In what follows we will prove the correctness of the main claim.

\item[{\em Proof of Main Claim:}] 

\item[A] Suppose $\dsigthree=\codsigthree$. 
Let $\lsigthree$ be a many-one complete language for $\sigmathree$.
It is not hard to verify that 
$\ldsigthree=\{\pair{x,y}\condition x\in \lsigthree \wedge 
y\not\in\lsigthree\}$
is a many-one complete language for $\dsigthree$. 
According to our assumption, $\dsigthree=\codsigthree$, there is a 
polynomial-time computable function $h$ reducing $\ldsigthree$ to 
$\ldsigthreebar$, 
i.e. for all $x_1,x_2 \in \sigmastar$,
$$\pair{x_1,x_2}\in \ldsigthree \iff h(\pair{x_1,x_2})\in \ldsigthreebar.$$
Let $h'$ and $h''$ be the polynomial-time computable functions such that 
for all $x_1,x_2 \in \sigmastar$,   
$h(\pair{x_1,x_2})=\pair{h'(\pair{x_1,x_2}),h''(\pair{x_1,x_2})}$. 
Hence
$$x_1 \in \lsigthree \wedge x_2 \not\in \lsigthree \iff
h'(\pair{x_1,x_2}) \not\in \lsigthree \vee h''(\pair{x_1,x_2}) 
\in \lsigthree.$$
The easy-hard method is based on the fact that $h$ is a many-one reduction 
from a conjunction to a disjunction.

\item[B] The string $x_2$ is said to be {\em easy\/} if and only 
if $(\exists x_1~ |x_1|=|x_2|)[h''(\pair{x_1,x_2}) \in \lsigthree]$. 
Clearly, if $x_2$ is easy then $x_2 \not\in \lsigthree$. 
But note that checking whether a particular string is easy can be done with a 
$\sigmathree$ algorithm.

$x_2$ is said to be {\em hard\/} if and only if 
$x_2 \not\in \lsigthree$ and 
$(\forall x_1~ |x_1|=|x_2|)[h''(\pair{x_1,x_2}) \not\in \lsigthree]$. 
Hence, if $x_2$ is a hard string we have for all $x_1$, $|x_1|=|x_2|$,
$$x_1\in \lsigthree\iff h'(\pair{x_1,x_2})\not\in\lsigthree.$$

Note that the strings in $\lsigthreebar$ divide into easy and hard strings.

\item[C] Define the set $S'=\{\omega\condition \omega 
\mbox{ is the lexicographically smallest hard string of length $|\omega|$}\}$ 
and
the set $S$ of marked prefixes of $S'$, 
$S=\{y\#^i\condition i\geq 0 \wedge (\exists v~|v|=i)
[yv \in S']\}$. 
Note that $S$ is sparse.

\item[D] {\em Claim~D:} $\pithree \subseteq (\sigmathree)^S$.

We will prove the above claim by giving a $(\sigmathree)^S$ algorithm for 
$\lsigthreebar$:
\begin{enumerate}
\item On input $x$, $|x|=n$, check whether $S^{=n}$ is empty or not. 
This can be done by querying $0\#^{n-1} \in S$ and 
$1\#^{n-1} \in S$. 
Obviously, $S^{=n}=\emptyset$ if and only if both queries are answered 
``no.''
\item  
If $S^{=n}=\emptyset$ then there exists no hard string of length $n$. 
Hence, $x \in \lsigthreebar$ if and only if $x$ is easy. 
Thus, guess $x_1$, $|x_1|=n$, compute 
$h(\pair{x_1,x})$, and accept 
if and only if $h''(\pair{x_1,x}) \in \lsigthree$.
\item 
If $S^{=n}\not= \emptyset$ then there exists a hard string of length $n$. 
Retrieve the only string not 
containing $\#$ 
(recall that this is the lexicographically smallest hard string of length $n$) 
from $S^{=n}$, call it $\omega$, with adaptive queries to $S^{=n}$.
Compute $h(\pair{x,\omega})$ and accept if and only if 
$h'(\pair{x,\omega})\in \lsigthree$.
\end{enumerate}

According to {\bf B}, this algorithm is correct.

\item[E] By a result of Yap~\cite{yap:j:advice}, 
$\pithree \subseteq (\sigmathree)^S$ for a sparse set $S$ implies 
$\sigmafive=\Pi_5^{\littlep}$ and hence 
$\ph=\sigmafive$.

\item[\em End of Proof of Main Claim]
\end{description}
\qed

\paragraph{2) Wagner 1987~\cite{wag:t:n-o-q-87version}}

\begin{theorem}\label{tt:wag1}
If $\psigthreeone=\psigthreetwo$ then $\ph=\p^{\sigmafour}$. 
\end{theorem}

\proof
\begin{description}
\item[{\em Main Claim}] {\em If $\dsigthree=\codsigthree$ then $\ph= 
\p^{\sigmafour}$.}

Clearly, the theorem follows as an immediate corollary from the above claim. 
We now prove the main claim.

\item[{\em Proof of Main Claim:}] 

\item[A and B] As in the proof of Theorem~\ref{tt:kad} (Kadin 1987).

\item[C] Let $\lsigthree$, $\lsigfour$, and $\lsigfive$ be 
many-one complete languages 
for $\sigmathree$, $\sigmafour$, and $\sigmafive$, respectively, 
$p_4$ and $p_5$ be polynomials such that
$$\lsigfour=\{x\condition (\exists y~|y|\leq p_4(|x|))
[\pair{x,y}\not\in \lsigthree]\}$$
and 
$$\lsigfive=\{x\condition (\exists y~|y|\leq p_5(|x|))
[\pair{x,y}\not\in \lsigfour]\}.$$

\item[D] For all $x \in \sigmastar$ let 
$$f(x)=
\left\{
\begin{array}{l@{\qquad}p{8cm}}
1 & if there exists a hard string of length $|x|$\\
0 & if there exists no hard string of length $|x|$.
\end{array}
\right.
$$
It is not hard to see that $f \in {\rm FP}^{\sigmafour[1]}$, where 
${\rm FP}^{\sigmafour[j]}$ is defined similar to $\p^{\sigmafour[j]}$ with 
the modification that the base $\p$ machine computes a function instead of 
accepting a language.
Note that 
$f(x)$ is equal for all equal-length strings $x$.

We call $\pair{1^l,\#}$ a hard pair if and only if 
$f(1^l)=0$.
$\pair{1^l,y}$, $y \in \sigmastar$, is called a hard pair 
if and only if $y$ is a hard string of length $l$.

\item[E] One hard pair suffices to provide a reduction from 
$\lsigthreebar$ to a $\sigmathree$ language. 

\item[\phantom{E}] {\em Claim~E: There exists a set $A \in \sigmathree$ such that 
for all $x \in \sigmastar$, 
if $\pair{1^{|x|},\omega}$ is a hard pair then 
$$ x\not\in \lsigthree \iff \pair{x,\omega} \in A.$$}

Let $x \in \sigmastar$. 
Let $\pair{1^{|x|},\omega}$ be a hard pair 
(note that this implies $\omega \in \sigmastar \cup \{\#\}$).
Suppose $f(1^{|x|})=0$. 
Hence $\omega=\#$ and 
for every string $y$ such that $|y|=|x|$, $y \not\in \lsigthree$ 
if and only if $y$ is easy. 
This holds in particular for $x$ itself.
According to {\bf B} we thus have 
$$x \not\in \lsigthree \iff (\exists x_1~|x_1|=|x|)
[h''(\pair{x_1,x}) \in \lsigthree].$$
Now suppose that $f(1^{|x|})=1$, hence $\omega$ is a hard string of length 
$|x|$.
According to {\bf B} we obtain  
$$x\in \lsigthree \iff h'(\pair{x,\omega}) \not\in\lsigthree.$$

We define 
$A=\{\pair{x,\omega}\condition (\omega=\# \wedge (\exists x_1~|x_1|=|x|)
[h''(\pair{x_1,x}) \in \lsigthree]) \vee
(\omega \in \sigmastar \wedge h'(\pair{x,\omega}) \in\lsigthree)\}$. 
It is not hard to verify that $A \in \sigmathree$ and 
that $A$ satisfies Claim~E.

\item[F] Applying Claim~E, a series of hard pairs of growing length 
gives a reduction from $\lsigfour$ to a $\sigmathree$ language.  

\item[\phantom{F}] {\em Claim~F: There exist a set $B \in \sigmathree$ and 
a polynomial $q$ such that for all $x \in \sigmastar$, 
if for all $0\leq i\leq q(|x|)$, $\pair{1^i,\omega_i}$ 
is a hard pair then 
$$ x\in \lsigfour \iff \pair{x,\omega_0,\omega_1,\dots ,\omega_{q(|x|)}} 
\in B.$$}

Let $x \in \sigmastar$. 
By definition of $\lsigfour$ we have
$$x \in \lsigfour \iff 
(\exists y~|y|\leq p_4(|x|))[\pair{x,y} \not\in \lsigthree].$$
According to Claim~E there exists a set $A \in \sigmathree$ such that 
for all $y\in \sigmastar$ if 
$\pair{1^{|\pair{x,y}|},\omega}$ 
is a hard pair then
$$\pair{x,y}\not\in \lsigthree \iff \pair{\pair{x,y},\omega} 
\in A.$$
Hence if for all $0\leq i\leq s(p_4(|x|))$, 
$\pair{1^i,\omega_i}$ 
($\omega_i \in \sigmastar \cup \{\#\}$)  
is a hard pair then
$$x \in \lsigfour \iff (\exists y~|y|\leq p_4(|x|))
[ \pair{\pair{x,y},\omega_{|\pair{x,y}|}} 
\in A].$$
Let $q$ be a polynomial such that $q(n)\geq s(p_4(n))$ for all $n$. 
Define 
$$B=\{\pair{x,\omega_0,\omega_1,\dots ,\omega_{q(|x|)}} \condition 
(\exists y~|y|\leq p_4(|x|))
[ \pair{\pair{x,y},\omega_{|\pair{x,y}|}} 
\in A]\}$$
and note that $B \in \sigmathree$.
This proves Claim~F.

\item[G] Taking the result of Claim~F one step further, hard pairs of 
growing length provide a reduction from $\lsigfive$ 
to a $\sigmafour$ language.

\item[\phantom{G}]{\em Claim~G: There exist a set $D \in \sigmafour$ and 
a polynomial $p$ such that for all $x \in \sigmastar$, 
if for all $0\leq i\leq p(|x|)$, $\pair{1^i,\omega_i}$ 
is a hard pair then 
$$ x\in \lsigfive \iff \pair{x,\omega_0,\omega_1,\dots ,\omega_{p(|x|)}} 
\in D.$$}

The proof is similar to the proof of Claim~F.
Let $x \in \sigmastar$.
We have 
$$x \in \lsigfive \iff 
(\exists y~|y|\leq p_5(|x|))[\pair{x,y} \not\in \lsigfour].$$
According to Claim~F there exist a set $B \in \sigmathree$ and 
a polynomial $q$ such that for all 
$y$, $|y|\leq p_5(|x|)$, 
if for all $0\leq i\leq q(s(p_5(|x|)))$, $\pair{1^i,\omega_i}$ 
is a hard pair, then
$$ \pair{x,y} \in \lsigfour \iff 
\pair{\pair{x,y},\omega_0,\omega_1,\dots ,\omega_{q(|\pair{x,y}|)}} \in B,$$
and hence 
$$x\in \lsigfive \iff 
(\exists y~|y|\leq p_5(|x|))
[\pair{\pair{x,y},\omega_0,\omega_1,\dots ,\omega_{q(|\pair{x,y}|)}} 
\not\in B].$$
Let $p$ be a polynomial such that $p(n)\geq q(s(p_5(n)))$ for all $n$.
Define 
$$D=\{\pair{x,\omega_0,\omega_1,\dots ,\omega_{p(|x|)}} \condition 
(\exists y~|y|\leq p_5(|x|))
[\pair{\pair{x,y},\omega_0,\omega_1,\dots ,\omega_{q(|\pair{x,y}|)}} 
\not\in B]\}.$$
Clearly, $D \in \sigmafour$.

\item[H] Since the hard pairs needed in Claim~G can be computed with queries 
to a $\sigmafour$ oracle we finally obtain

\item[\phantom{H}]{\em Claim~H: $\lsigfive \in \p^{\sigmafour}$.}

Let $D \in \sigmafour$ and $p$ be a polynomial as defined in {\bf G}.
We have the following $\p^{\sigmafour}$ algorithm for $\lsigfive$:

\begin{enumerate}
\item 
On input $x$, compute $f(1^0)$, $f(1^1)$, $f(1^2)$, \dots , $f(1^{p(|x|)})$. 
This can be done with $p(|x|)$ parallel queries to a $\sigmafour$ oracle 
since $f \in {\rm FP}^{\sigmafour[1]}$. 
\item 
\begin{enumerate} 
\item 
For all $0\leq i \leq p(|x|)$ such that $f(1^i)=0$ set $\omega_i=\#$.
\item 
For all $0\leq i \leq p(|x|)$ such that $f(1^i)=1$ guess 
$\omega_i$, $|\omega_i|=i$, and verify that $\omega_i$ is a hard string. 
Continue if this verification succeeds, otherwise reject. 
\item Verify that  
$\pair{x,\omega_0,\omega_1,\dots ,\omega_{p(|x|)}} \in D$.
\end{enumerate}
Note that all this can be done with a single $\sigmafour$ oracle query
when handing $x$, $f(1^0)$, $f(1^1)$, $f(1^2)$, \dots , and $f(1^{p(|x|)})$ 
over to the oracle.
\item 
Accept if and only if 
the query under 2. returns ``yes.''
\end{enumerate}

It is not hard to verify that the above algorithm in light of Claim~G 
proves the claim.
Note that the oracles queried in steps 1 and 2 might be different. 
Since $\sigmafour$ is closed under disjoint union this 
can easily be avoided by using the disjoint union of the oracles and 
modifying the oracle queries in such a way that they are made to the 
correct part of the disjoint union. 

\item[I] Since $\lsigfive$ is complete for $\sigmafive$ we conclude 
$\sigmafive=\p^{\sigmafour}$ and thus $\ph = \p^{\sigmafour}$.

\item[\em End of Proof of Main Claim]
\end{description}
\qed

\paragraph{3) Wagner 1989~\cite{wag:t:n-o-q-89version}}

\begin{theorem}\label{tt:wag2}
If $\psigthreeone=\psigthreetwo$ then $\ph=\p^{\sigmafour[3]}$.
\end{theorem}

\proof
\begin{description}
\item[{\em Main Claim}] As in the proof of Theorem~\ref{tt:wag1} 
(Wagner 1987), 
but $\p^{\sigmafour}$ replaced by 
$\p^{\sigmafour[3]}$.

\item[\em Proof of Main Claim:]
The major difference to the proof of Theorem~\ref{tt:wag1} lies in {\bf B} of 
the upcoming proof, 
a modified definition of easy and hard strings. 
Note that a straightforward adaption of the proof of Theorem~\ref{tt:wag1} 
to this new definition would suffice to prove Theorem~\ref{tt:wag2}. 
However, Wagner used a slightly different approach and obtained stronger 
intermediate results than in~\cite{wag:t:n-o-q-87version}. 
Though those stronger intermediate results do not lead to a better  
overall result, they appear in the papers of 
Chang and Kadin~\cite{cha-kad:toutdated:bh,cha-kad:j:closer} and 
Beigel, Chang, and 
Ogihara~\cite{bei-cha-ogi:tOUT:boolean,bei-cha-ogi:j:difference-hierarchies} 
again and play an important role there. 

\item[A] As in the proof of Theorem~\ref{tt:kad} (Kadin 1987).

\item[B] Let $l$ be an integer. 
The string $x_2$ is said to be {\em easy for length $l$\/} 
if and only if $|x_2|\leq l$ and 
$(\exists x_1~ |x_1|\leq l)[h''(\pair{x_1,x_2}) \in \lsigthree]$. 
Clearly, if $x_2$ is easy for length $l$ then $x_2 \not\in \lsigthree$. 

$x_2$ is said to be {\em hard for length $l$\/} if and only if 
$|x_2|\leq l$,  
$x_2 \not\in \lsigthree$, and 
$(\forall x_1~ |x_1|\leq l)[h''(\pair{x_1,x_2}) \not\in \lsigthree]$. 
Hence, if $x_2$ is a hard string for length $l$ we have for all $x_1$, 
$|x_1|\leq l$,
$$x_1\in \lsigthree\iff h'(\pair{x_1,x_2})\not\in\lsigthree.$$

Note that the strings in ${(\lsigthreebar)}^{\leq l}$ divide 
into easy and hard strings for length $l$.

\item[C] As in the proof of Theorem~\ref{tt:wag1} (Wagner 1987).

\item[D] 
For all $x \in \sigmastar$ let 
$$f(x)=
\left\{
\begin{array}{l@{\qquad}p{8cm}}
1 & if there exists a hard string {\em for\/} length $|x|$\\
0 & if there exists no hard string {\em for\/} length $|x|$.
\end{array}
\right.
$$
Note, $f \in {\rm FP}^{\sigmafour[1]}$ and 
$f(x)$ is equal for all equal-length strings $x$.

We call $\pair{1^l,\#}$ a hard pair if and only if 
$f(1^l)=0$.
$\pair{1^l,y}$, $y \in \sigmastar$, is called a hard pair 
if and only if $y$ is a hard string for length $l$.

\item[E] 
Similar to {\bf E} in the proof of Theorem~\ref{tt:wag1} (Wagner 1987) 
(with the obvious adaptions due to the changed definition of easy and hard 
strings)
one hard pair gives a reduction from $\lsigthreebar$ to a 
$\sigmathree$ language.

\item[\phantom{E}]
{\em Claim~E: There exists a set $A \in \sigmathree$ such that 
for all $x \in \sigmastar$ and all $l \geq |x|$, 
if $\pair{1^l,\omega}$ 
is a hard pair then 
$$ x\not\in \lsigthree \iff \pair{x,1^l,\omega} \in A.$$}

Let $x \in \sigmastar$ and $l \geq |x|$. 
Suppose that  $\pair{1^l,\omega}$ is a hard pair, 
hence $\omega \in \sigmastar \cup \{\#\}$.

If $f(1^l)=0$ then $\omega=\#$ and for every string $y$, 
$|y|\leq l$, $y \in \lsigthreebar$ 
if and only if $y$ is easy for length $l$. 
This holds in particular for $x$ itself.
According to {\bf B} we thus have 
$$x \not\in \lsigthree \iff (\exists x_1~|x_1|\leq l)
[h''(\pair{x_1,x}) \in \lsigthree].$$
If $f(1^l)=1$ then  $\omega$ is a hard string for length $l$.
According to {\bf B} we obtain  
$$x\in \lsigthree \iff h'(\pair{x,\omega}) \not\in\lsigthree.$$

We define 
$A=\{\pair{x,1^l,\omega}\condition (\omega=\# \wedge (\exists x_1~|x_1|\leq l)
[h''(\pair{x_1,x}) \in \lsigthree]) \vee
(\omega \in \sigmastar \wedge h'(\pair{x,\omega}) \in\lsigthree)\}$. 
It is not hard to verify that $A \in \sigmathree$.
This completes the proof of Claim~E.

\item[F] 
In contrast to {\bf F} in the proof of Theorem~\ref{tt:wag1} (Wagner 1987), 
the new  definition of easy and hard strings yields that 
{\em one\/} hard pair for sufficiently large length suffices to reduce 
$\lsigfour$ to a $\sigmathree$ language. 

\item[\phantom{F}]
{\em Claim~F: There exist a set $B \in \sigmathree$ and 
a polynomial $q$ such that for all $x \in \sigmastar$ and all $l \geq q(|x|)$, 
if $\pair{1^l,\omega}$ is a hard pair then 
$$ x\in \lsigfour \iff \pair{x,1^l,\omega} \in B.$$}

Let $x \in \sigmastar$. 
By definition of $\lsigfour$ we have
$$x \in \lsigfour \iff 
(\exists y~|y|\leq p_4(|x|))[\pair{x,y} \not\in \lsigthree].$$
According to Claim~E there exists a set $A \in \sigmathree$ such that  
if $\pair{1^l,\omega}$, $l \geq s(p_4(|x|))$, 
is a hard pair then for all $y$, $|y|\leq p_4(|x|)$,
$$\pair{x,y}\not\in \lsigthree \iff \pair{\pair{x,y},1^l,\omega} 
\in A,$$
and hence
$$x \in \lsigfour \iff (\exists y~|y|\leq p_4(|x|))
[ \pair{\pair{x,y},1^l,\omega} 
\in A].$$
Let $q$ be a polynomial such that $q(n)\geq s(p_4(n))$ for all $n$. 
Define 
$B=\{\pair{x,1^l,\omega} \condition (\exists y~|y|\leq p_4(|x|))
[ \pair{\pair{x,y},1^l,\omega} \in A]\}$
and note that $B \in \sigmathree$.
This proves Claim~F.

\item[G] 
Applying Claim~F twice provides a reduction from $\lsigfive$ 
to a $\sigmathree$ language requiring two hard pairs.

\item[\phantom{G}]
{\em Claim~G: There exist a set $C \in \sigmathree$ and 
polynomials $q_1$ and $q_2$ such that for all $x \in \sigmastar$, 
if $\pair{1^{q_1(|x|)},\omega_1}$ and 
$\pair{1^{q_2(|x|)},\omega_2}$ are hard pairs then
$$ x\in \lsigfive \iff \pair{x,\omega_1,\omega_2} \in C.$$}

Let $x \in \sigmastar$.
We have 
$$x \in \lsigfive \iff 
(\exists y~|y|\leq p_5(|x|))[\pair{x,y} \not\in \lsigfour].$$
According to Claim~F there exist a set $B \in \sigmathree$ and a 
polynomial $q$ such that if $\pair{1^l,\omega_1}$, 
$l  \geq q(s(p_5(|x|)))$, is a hard pair then 
$$
x \in \lsigfive \iff 
(\exists y~|y|\leq p_5(|x|))[\pair{\pair{x,y},1^l,\omega_1} \not\in B].
$$
Let $q_1$ be a polynomial such that $q_1(n)\geq q(s(p_5(n)))$ for all $n$.
Define 
$$
D=\{\pair{x,1^l,\omega_1}\condition 
(\exists y~|y|\leq p_5(|x|))[\pair{\pair{x,y},1^l,\omega_1} 
\not\in B]\}.$$
Note that $D\in \sigmafour$ and let $g$ be a many-one reduction from $D$ 
to $\lsigfour$. 
Let $\wh{q}$ be a polynomial such that for all $z \in \sigmastar$, 
$|g(z)|\leq \wh{q}(|z|)$. 
Hence we have that if $\pair{1^{q_1(|x|)},\omega_1}$ is a hard pair then 
$$
x \in \lsigfive \iff g(\pair{x,1^{q_1(|x|)},\omega_1})\in \lsigfour.
$$
Applying Claim~F again we obtain that if 
$\pair{1^l,\omega_2}$, $l \geq q(\wh{q}(s(q_1(|x|))))$, 
is a hard pair and $|\omega_1|\leq q_1(|x|)$,
$$
g(\pair{x,1^{q_1(|x|)},\omega_1})\in \lsigfour \iff
\pair{g(\pair{x,1^{q_1(|x|)},\omega_1}),1^l,\omega_2} \in B.
$$
All together, if 
$\pair{1^{q_1(|x|)},\omega_1}$, 
and 
$\pair{1^l,\omega_2}$, $l\geq q(\wh{q}(s(q_1(|x|))))$, 
are hard pairs then
$$
x \in \lsigfive \iff 
\pair{g(\pair{x,1^{q_1(|x|)},\omega_1}),1^l,\omega_2} \in B.
$$
Let $q_2$ be a polynomial such that $q_2(n)\geq q(\wh{q}(s(q_1(n))))$ 
for all $n$.
Define $C=\{\pair{x,\omega_1,\omega_2}\condition 
\pair{g(\pair{x,1^{q_1(|x|)},\omega_1}),1^{q_2(|x|)},\omega_2} \in B\}$.
This completes the proof of Claim~G.
\footnote{
The reader will observe that in light of the algorithm given in {\bf H} below  
a reduction of $\lsigfive$ to the $\sigmafour$ language 
(as it was done in the proof of Theorem~\ref{tt:wag1} (Wagner 1987))
would suffice. 
This is implicitly done by the set $D$ in the above proof of Claim~G.
Since this would clearly require only one hard pair 
one could similarly to {\bf H} derive even 
$\lsigfive \in  \p^{\sigmafour[2]}$. 
However, the possibility of reducing $\lsigfive$ to a $\sigmathree$ 
language first appeared in~\cite{wag:t:n-o-q-89version} and was crucially used 
in~\cite{cha-kad:toutdated:bh,cha-kad:j:closer} 
and~\cite{bei-cha-ogi:tOUT:boolean,bei-cha-ogi:j:difference-hierarchies}.
}

\item[H] 
In contrast to {\bf H} of the proof of Theorem~\ref{tt:wag1} (Wagner 1987), 
only two values of $f$ have to be computed in light of Claim~G. 

\item[\phantom{H}]
{\em Claim~H: $\lsigfive \in \p^{\sigmafour[3]}$.}

Let $C \in \sigmathree$ and $q_1$ and $q_2$ be polynomials as defined 
in {\bf G}.
We give a $\p^{\sigmafour[3]}$ algorithm for $\lsigfive$.
\begin{enumerate}
\item 
On input $x$ compute $f(1^{q_1(|x|)})$ and $f(1^{q_2(|x|)})$. 
This amounts for two $\sigmafour$ queries.
\item 
\begin{enumerate}
\item 
If $f(1^{q_1(|x|)})=0$ set $\omega_1=\#$. 
If $f(1^{q_1(|x|)})=1$ guess $\omega_1$, $|\omega_1|\leq q_1(|x|)$, 
and verify that $\omega_1$ is a hard string for length $q_1(|x|)$. 
Continue if this verification succeeds, otherwise reject.
\item 
If $f(1^{q_2(|x|)})=0$ set $\omega_2=\#$. 
If $f(1^{q_2(|x|)})=1$ guess $\omega_2$, $|\omega_2|\leq q_2(|x|)$, 
and verify that $\omega_2$ is a hard string for length $q_2(|x|)$. 
Continue if this verification succeeds, otherwise reject.
\item Verify that  
$\pair{x,\omega_1,\omega_2}\in C$.
\end{enumerate}
Note that all this can be done with a single $\sigmafour$ oracle query when 
handing $x$, $f(1^{q_1(|x|)})$, and $f(1^{q_2(|x|)})$ over to the oracle.
\item 
Accept if and only if 
the query under 2. returns ``yes.''
\end{enumerate}

The correctness of this algorithm is obvious, in particular recall from 
{\bf H} of the proof of Theorem~\ref{tt:wag1} (Wagner 1987) that the use 
of different 
$\sigmafour$ oracles does no harm to the algorithm.

\item[I]
Since $\lsigfive$ is complete for $\sigmafive$ we have 
$\sigmafive =\p^{\sigmafour[3]}$ and hence 
$\ph = \p^{\sigmafour[3]}$.

\item[\em End of Proof of Main Claim]

\end{description}
\qed

\paragraph{4) Chang/Kadin 1989~\cite{cha-kad:toutdated:bh,cha-kad:j:closer}}

\begin{theorem}\label{tt:ck}
If $\psigthreeone=\psigthreetwo$ then $\ph= {\rm D}\cdot\sigmafour$.
\end{theorem}

\proof
\begin{description}
\item[\em Main Claim] {\em If $\dsigthree=\codsigthree$ then 
$\ph = {\rm D}\cdot\sigmafour$.}

Obviously, it suffices to prove the main claim.

\item[\em Proof of Main Claim:]

\item[A] As in the proof of Theorem~\ref{tt:kad} (Kadin 1987).

\item[B,C,D,E,F, and G] As in the proof of Theorem~\ref{tt:wag2} (Wagner 1989).

\item[H] In contrast to {\bf G} one hard pair suffices to reduce a 
$\sigmafive$ complete language to a $\sigmafour$ language.

\item[\phantom{H}]
{\em Claim~H: There exist a set $D_1 \in \sigmafour$ 
and a polynomial $p_1$ such that for all $x\in \sigmastar$ and 
all $l \geq p_1(|x|)$, 
if $\pair{1^l,\omega}$ is a hard pair then 
$$x \in \lsigfive \iff \pair{x,1^l,\omega} \in D_1.$$}

Claim~H is the analogue of 
{\bf G} in the proof of Theorem~\ref{tt:wag1} (Wagner 1987) 
with the modifications induced by the different hard strings definition 
and 
is implicitly contained in {\bf G} of the proof of  
Theorem~\ref{tt:wag2} (Wagner 1989). 
In particular, setting $p_1=q_1$ and $D_1=D$, where $q_1$ and $D$ are as 
defined in {\bf G} of the proof of Theorem~\ref{tt:wag2} (Wagner 1989) 
proves the claim.

\item[I] Define $S=\{1^l\condition f(1^l)=1\}$. 
Though $f \in {\rm FP}^{\sigmafour[1]}$, testing whether $f(1^l)=1$ can be 
done with a $\sigmafour$ algorithm, as it is just testing whether there 
exists a hard string for length $l$. 
So $S\in \sigmafour$.

\item[\phantom{I}]
{\em Claim~I: There exist a set $T \in \sigmathree$ 
and a polynomial $p_t$ such that for all $l\in \natnum$ 
and all $l' \geq p_t(l)$, 
if $\pair{1^{l'},\omega}$ is a hard pair then
$$
1^l \in S \iff \pair{1^l,1^{l'},\omega} 
\in T.$$}

The claim follows immediately from {\bf F}.

\item[J] 
The above Claim~I turns into the key tool to reduce a $\pifive$ complete 
language to a $\sigmafour$ language with the help of just one hard pair.

\item[\phantom{J}]
{\em Claim~J: There exist a set $D_2 \in \sigmafour$ 
and a polynomial $p_2$ such that for all $x\in \sigmastar$ and all 
$l \geq p_2(|x|)$, 
if $\pair{1^l,\omega}$ is a hard pair then
$$x \not\in \lsigfive \iff \pair{x,1^l,\omega} \in D_2.$$}

The reader will soon observe that Claim~J is the key trick in the current 
proof.

Let $p_t$, $q_1$, and  $q_2$ be polynomials and $C \in \sigmathree$ 
as defined in {\bf G} and {\bf I}.
Let $p_2$ be a polynomial such that 
$p_2(n) \geq p_t(q_2(n))$ for all $n$. 
$D_2$ is defined by the following $\sigmafour$ algorithm.
\begin{enumerate}
\item 
On input $\pair{x,1^l,\omega}$, compute $q_1(|x|)$ and $q_2(|x|)$.
\item 
Assuming that $\pair{1^l,\omega}$ is a hard pair and $l\geq p_t(q_1(|x|))$ 
we determine $f(1^{q_1(|x|)})$ by applying Claim~I. 
This is done as follows: 
Test using a $\sigmathree$ oracle query whether 
$\pair{1^{q_1(|x|)},1^l,\omega}\in T$.
Set $j_1=1$ if this is the case, otherwise $j_1=0$.
\item 
Assuming that $\pair{1^l,\omega}$ is a hard pair and $l\geq p_t(q_2(|x|))$ 
we determine $f(1^{q_2(|x|)})$ by applying Claim~I. 
This is done similar to step 2: 
Test using a $\sigmathree$ oracle query whether 
$\pair{1^{q_2(|x|)},1^l,\omega}\in T$.
Set $j_2=1$ if this is the case, otherwise $j_2=0$.
\item 
If $j_1=1$ guess a string $\omega_1$, $|\omega_1|\leq q_1(|x|)$, verify that 
$\omega_1$ is hard for length $q_1(|x|)$, continue if this is the case, and 
reject otherwise. 
If $j_1=0$ set $\omega_1=\#$.
\item 
If $j_2=1$ guess a string $\omega_2$, $|\omega_2|\leq q_2(|x|)$, verify that 
$\omega_2$ is hard for length $q_2(|x|)$, continue if this is the case, and 
reject otherwise. 
If $j_2=0$ set $\omega_2=\#$.
\item 
Assuming that $\pair{1^{q_1(|x|)},\omega_1}$ and 
$\pair{1^{q_2(|x|)},\omega_2}$ are hard pairs we determine  
whether $x \not\in \lsigfive$ using Claim~G.
In other words, 
accept if and only if $\pair{x,\omega_1,\omega_2} \not\in C$. 
\end{enumerate}

Observe that if $\pair{1^l,\omega}$ is a hard pair and 
$l\geq p_t(q_1(|x|))$, step 2 indeed yields $j_1=f(1^{q_1(|x|)})$ 
according to {\bf I}. 
Similarly, if $\pair{1^l,\omega}$ is a hard pair and 
$l\geq p_t(q_2(|x|))$, step 3 correctly determines $j_2=f(1^{q_2(|x|)})$ 
according to {\bf I}. 
Furthermore, if $\pair{1^{q_1(|x|)},\omega_1}$ and 
$\pair{1^{q_2(|x|)},\omega_2}$ are hard pairs then we accept in step 6 
if and only if $x \notin\lsigfive$. 
But, if steps 2 and 3 yield $j_1=f(1^{q_1(|x|)})$ and $j_2=f(1^{q_2(|x|)})$, 
respectively, the algorithm indeed determines hard pairs in steps 4 and 5 and 
hence the algorithm correctly accepts in step 6.

Overall, the correctness of the above algorithm stands and falls with 
correctness of steps 2 and 3.
Hence setting $p_2(n) \geq p_t(q_2(n))$ for all $n$ 
proves the claim 
(note that in light of our convention about polynomials 
$q_2(n)>q_1(n)$ for all $n$).

\item[K] 
Combining the results of Claims~H and J while exploiting the difference 
structure of ${\rm D}\cdot{\sigmafour}$ yields

\item[\phantom{K}]
{\em Claim~K: $\lsigfive \in {\rm D}\cdot{\sigmafour}$.}
 
Let the sets $D_1,D_2 \in \sigmafour$ and the polynomials $p_1$ and $p_2$ 
be as defined in {\bf H} and {\bf J}.
Let $p$ be a polynomial such that 
$p(n)\geq \max\{p_1(n),p_2(n)\}$ for all $n$.
Define 
$$
E_1=\{x \condition \pair{x,1^{p(|x|)},\#}\in D_1\},
$$ 
$$
E_2=\{x \condition (\exists \omega\in \sigmastar)[\omega 
\mbox{ is a hard string for length } p(|x|) \mbox{ and } 
\pair{x,1^{p(|x|)},\omega} \in D_1\},
$$ 
and 
$$
E_3=\{x \condition (\exists \omega\in \sigmastar)[\omega 
\mbox{ is a hard string for length } p(|x|) \mbox{ and } 
\pair{x,1^{p(|x|)},\omega} \in D_2\}.
$$ 
Clearly, $E_1,E_2,E_3 \in \sigmafour$.
Since $\sigmafour$ is closed under union we also have 
$E_1\cup E_2\in \sigmafour$.
Hence $(E_1\cup E_2)-E_3 \in {\rm D}\cdot{\sigmafour}$.
We show $\lsigfive=(E_1\cup E_2)-E_3$. 
To see this  consider the following case distinction.
Let $x \in \sigmastar$.
\begin{description}
\item[Case 1] 
$f(1^{p(|x|)})=0$.\\
Hence $x \not\in E_2$ and $x \not\in E_3$.
Furthermore, $\pair{1^{p(|x|)},\#}$ is a hard pair and hence according to 
Claim~H,
\begin{eqnarray*}
x\in \lsigfive &\iff& \pair{x,1^{p(|x|)},\#} \in D_1\\
&\iff& x \in E_1\\
&\iff& x \in (E_1\cup E_2)-E_3.
\end{eqnarray*}

\item[Case 2] 
$f(1^{p(|x|)})=1$.\\
Hence there exist hard strings for length 
$p(|x|)$. 
If $x \in \lsigfive$ then clearly $\pair{x,1^{p(|x|)},\omega} \in D_1$ 
and $\pair{x,1^{p(|x|)},\omega} \not\in D_2$ 
for all hard strings $\omega$ for length $p(|x|)$,
according to Claims~H and J. 
Hence $x \in (E_1 \cup E_2)-E_3$.
If $x \not\in \lsigfive$ then  
$\pair{x,1^{p(|x|)},\omega} \not\in D_1$ 
and $\pair{x,1^{p(|x|)},\omega} \in D_2$ 
for all hard strings $\omega$ for length $p(|x|)$,
according to Claims~H and J.
Independent of whether $x \in E_1$ or $x \not\in E_1$ we have 
$x \not\in (E_1 \cup E_2)-E_3$.
\end{description}

\item[L] 
We have shown $\lsigfive \in {\rm D}\cdot\sigmafour$ and thus 
$\sigmafive={\rm D}\cdot\sigmafour$ which immediately implies 
$\ph = {\rm D}\cdot\sigmafour$.

\item[\em End of Proof of Main Claim]

\end{description}
\qed

\paragraph{5) Beigel/Chang/Ogiwara 1991~\cite{bei-cha-ogi:tOUT:boolean,bei-cha-ogi:j:difference-hierarchies}}

\begin{theorem}\label{tt:bco}
If $\psigthreeone=\psigthreetwo$ then 
$\ph = \left(\p^{\np}_{1\hbox{-}{\rm tt}}\right)^{\sigmathree}$.
\end{theorem}

\proof
\begin{description}

\item[\em Main Claim] {\em If $\dsigthree=\codsigthree$ then 
$\ph = \left(\p^{\np}_{1\hbox{-}{\rm tt}}\right)^{\sigmathree}$,}

where $\left(\p^{\np}_{1\hbox{-}{\rm tt}}\right)^{\sigmathree}$ is 
the class of languages accepted by some DPTM making at most one query 
to a $\np^{\sigmathree}=\sigmafour$ oracle and 
polynomially many queries to a $\sigmathree$ oracle.

The theorem follows immediately from the main claim, which we will prove now. 

\item[\em Proof of Main Claim:]

\item[A] As in the proof of Theorem~\ref{tt:kad} (Kadin 1987). 

\item[B,C,D,E,F and G] As in the proof of Theorem~\ref{tt:wag2} (Wagner 1989). 

\item[H] Quite similar to {\bf H} of the proof of Theorem~\ref{tt:ck} 
(Chang/Kadin 1989) one hard pair suffices to reduce a 
$\p^{\sigmafour}$ language to a $\p^{\sigmathree}$ language.

\item[\phantom{H}]
{\em Claim~H: Let $L \in \p^{\sigmafour}$. 
There exist a set $D \in \p^{\sigmathree}$ and a polynomial $p$ 
such that for all $x \in \sigmastar$, 
if $\pair{1^{p(|x|)},\omega}$ is a hard pair then 
$$
x \in L \iff \pair{x,\omega} \in D.
$$}

The proof is a straightforward application of {\bf F}. 
Let $L\in \p^{\sigmafour}$, hence $L=L(N_1^{\lsigfour})$ for some 
DPTM $N_1$ running in time $\wh{p}$ for some polynomial $\wh{p}$. 
According to {\bf F} there exist a language $B \in \sigmathree$ 
and a polynomial $q$ such that for all $x \in \sigmastar$ 
and all $l\geq q(|x|)$, 
if $\pair{1^l,\omega}$ is a hard pair then 
$$x \in \lsigfour \iff \pair{x,1^l,\omega}\in B.$$

We use this to reduce $L$ to a 
$\p^{\sigmathree}$ language with the help of one hard pair. 
Let $p$ be a polynomial such that  $p(n) \geq  q(\wh{p}(n))$ for all 
$n$. 
Define the DPTM $N_2^B$ as follows:  
$N_2^B(\pair{x,\omega})$ simulates the work of $N_1^{\lsigfour}(x)$ 
but replaces every query $v$ to $\lsigfour$ by a  query 
$\pair{v,1^{p(|x|)},\omega}$ to $B$. 
Let $D=L(N_2^B)$. 
Clearly,  $D \in \p^{\sigmathree}$.
It is not hard to verify that this proves the claim.

\item[I] {\em Claim~I: $\lsigfive \in 
\left(\p^{\np}_{2\hbox{-}{\rm tt}}\right)^{\sigmathree}$.}

According to Claim~G there exist 
a language $C\in \sigmathree$ and polynomials $q_1$ and $q_2$ 
such that for all $x \in \sigmastar$,
if $\pair{1^{q_1(|x|)},\omega_1}$ and  
$\pair{1^{q_2(|x|)},\omega_2}$ are hard pairs then  
$$x \in \lsigfive  \iff  
\pair{x,\omega_1,\omega_2}\in C.$$

If no hard strings for length $q_1(|x|)$ and $q_2(|x|)$ exist then 
$x\in \lsigfive \iff \pair{x,\#,\#}\in C$. 
On the other hand, we can rely on $\chi_C(\pair{x,\#,\#})$ for determining 
$\chi_{\lsigfive}(x)$ if we know whether existing hard strings for 
length $q_1(|x|)$ and $q_2(|x|)$ provide 
$x\in \lsigfive \iff \pair{x,\#,\#}\not\in C$ or not. 
This approach is known as the mind-change technique.
It enables us to 
give a $\left(\p^{\np}_{2\hbox{-}{\rm tt}}\right)^{\sigmathree}$ 
algorithm for $\lsigfive$.
\begin{enumerate}
\item 
On input $x$, determine whether 
$\pair{x,\#,\#}\in C$.
This can be done with one query to $C$.
\item 
\nopagebreak%
\begin{enumerate}%
\nopagebreak%
\item Determine whether 
\begin{enumerate}%
\item there exists 
a hard string $\omega_1$ for length $q_1(|x|)$ such that 
$\chi_C(\pair{x,\omega_1,\#})\not=
\chi_C(\pair{x,\#,\#})$, or
\item there exists 
a hard string $\omega_2$ for length $q_2(|x|)$ such that 
$\chi_C(\pair{x,\#,\omega_2})\not=
\chi_C(\pair{x,\#,\#})$, or 
\item there exist two hard strings $\omega_1$ and $\omega_2$ for length 
$q_1(|x|)$ and $q_2(|x|)$, respectively, such that 
$\chi_C(\pair{x,\omega_1,\omega_2})\not=
\chi_C(\pair{x,\#,\#})$.
\end{enumerate}
\item  
Determine whether there exist
two hard strings $\omega_1$ and $\omega_2$ for length $q_1(|x|)$ 
and $q_2(|x|)$, respectively, such that either\nopagebreak%
\begin{enumerate}%
\item $\chi_C(\pair{x,\omega_1,\omega_2})\not=
\chi_C(\pair{x,\omega_1,\#})$ 
and 
$\chi_C(\pair{x,\omega_1,\#})\not=
\chi_C(\pair{x,\#,\#})$ or
\item $\chi_C(\pair{x,\omega_1,\omega_2})\not=
\chi_C(\pair{x,\#,\omega_2})$ 
and 
$\chi_C(\pair{x,\#,\omega_2})\not=
\chi_C(\pair{x,\#,\#})$. 
\end{enumerate}
\end{enumerate}
Note that all this can be done with two parallel queries to a $\sigmafour$ 
oracle, one query for (a) and one for (b).

\item
Accept if and only if the three queries from 1., 2.(a), and 2.(b) return in 
this order the answers ``yes,no,no,'' ``no,yes,no,'' or ``yes,yes,yes.''
\end{enumerate}
The correctness of this algorithm follows immediately from the construction. 
As already pointed out in {\bf H} in the proof of Theorem~\ref{tt:wag1} 
(Wagner 1987) the use of different 
$\sigmafour$ oracles in step 2 does not affect the correctness of 
our algorithm.
Note that step 2.(a) corresponds to checking whether the existence of hard 
strings causes at least one mind change, whereas step 2.(b) corresponds to 
determining whether the existence of hard strings causes two mind changes.

\item[J] In order to prove the {\bf Main Claim} one somehow has to show 
that  the $\left(\p^{\np}_{2\hbox{-}{\rm tt}}\right)^{\sigmathree}$ 
algorithm for $\lsigfive$ as given in {\bf I} can be improved to a 
$\left(\p^{\np}_{1\hbox{-}{\rm tt}}\right)^{\sigmathree}$ algorithm.
This is achieved by exploiting Claim~H.

\item[\phantom{J}]
{\em Claim~J: $\p^{\sigmafour}\subseteq 
\left(\p^{\np}_{1\hbox{-}{\rm tt}}\right)^{\sigmathree}$.}

Let $L\in \p^{\sigmafour}$.
Let $D \in \p^{\sigmathree}$ and $p$ be a polynomial as defined in 
{\bf H}. 
We describe a $\left(\p^{\np}_{1\hbox{-}{\rm tt}}\right)^{\sigmathree}$ 
algorithm for $L$ which uses the same idea as in {\bf I}.
\begin{enumerate}
\item 
On input $x$ determine whether $\pair{x,\#}\in D$. 
This can be done with the help of queries to a $\sigmathree$ oracle, since 
$D \in \p^{\sigmathree}$.
\item 
Check whether there exists 
a hard string $\omega$ for length $p(|x|)$ such that 
$\chi_{D}(\pair{x,\omega})\not=\chi_{D}(\pair{x,\#})$.
This can be done with one query to a $\sigmafour$ oracle.
\item Accept if and only if the two queries return different answers.
\end{enumerate}
The correctness of this algorithm follows immediately from the construction.
Note that step 2 corresponds to determining whether the existence of a hard 
string causes a mind change.

\item[K] Since 
$\sigmafive \subseteq 
\left(\p^{\np}_{2\hbox{-}{\rm tt}}\right)^{\sigmathree}$ 
(Claim~I), $\left(\p^{\np}_{2\hbox{-}{\rm tt}}\right)^{\sigmathree} 
\subseteq  \p^{\sigmafour}$, and 
$\p^{\sigmafour}\subseteq 
\left(\p^{\np}_{1\hbox{-}{\rm tt}}\right)^{\sigmathree}$ (Claim~J) 
we have proven 
$\sigmafive \subseteq \left(\p^{\np}_{1\hbox{-}{\rm tt}}\right)^{\sigmathree}$ 
and thus $\ph = \left(\p^{\np}_{1\hbox{-}{\rm tt}}\right)^{\sigmathree}$.

\item[\em End of Proof of Main Claim]

\end{description}
\qed

\paragraph{6) Hemaspaandra/Hemaspaandra/Hempel 1996~\cite{hem-hem-hem:tOutBydown-sep:upward-sep}}

\begin{theorem}\label{tt:hhh1}
If $\psigthreeone=\psigthreetwo$ then $\ph=\sigmathree$.
\end{theorem}

\proof
\begin{description}

\item[\em Main Claim] {\em If $\p^{(\p,\sigmathree)} = \p^{(\np,\sigmathree)}$ 
then  $\ph=\sigmathree$,}  

where $\p^{(\p,\sigmathree)}$ and $\p^{(\np,\sigmathree)}$ are 
the classes 
of languages that can be accepted by some DPTM making {\em in parallel\/} 
at most 
one query to a $\p$ or $\np$ oracle, respectively, and at most one query to a 
$\sigmathree$ oracle. 

Since $\psigthreeone \subseteq \p^{(\p,\sigmathree)} \subseteq 
\p^{(\np,\sigmathree)} \subseteq \psigthreetwo$ 
the theorem follows immediately from the above claim.
Thus, it remains to prove the main claim.

\item[\em Proof of Main Claim:]

\item[A] Suppose $\p^{(\p,\sigmathree)} = \p^{(\np,\sigmathree)}$.
Let $\lp$ and $\lnpone$ be many-one complete languages for $\p^{\p[1]}=\p$ and 
$\p^{\np[1]}$, respectively. 
Let $\lsigthree$ be a $\sigmathree$ complete language.
In order to prove the {\bf Main Claim} it suffices to give a
$\sigmathree$ algorithm for $\lsigthreebar$.

\item[B] Define for any two sets $A$ and $B$,
$A\deltatilde B=\{\pair{x,y}\condition x\in A \iff y\not\in B\}$.

\item[\phantom{B}]  
{\em Claim~B: $\lp\deltatilde\lsigthree$ and $\lnpone\deltatilde\lsigthree$ 
are many-one complete languages for $\p^{(\p,\sigmathree)}$ and 
$\p^{(\np,\sigmathree)}$, respectively.}

We will only show the claim for $\lnpone\deltatilde\lsigthree$. 
Obviously, $\lnpone\deltatilde\lsigthree \in \p^{(\np,\sigmathree)}$.
Let $L$ be an arbitrary language from $\p^{(\np,\sigmathree)}$. 
Without loss of generality let $L$ be accepted by a DPTM $N$ 
making, on every input $x$,  in parallel exactly one query $x_A$ to $A$ 
and one query $x_B$ to $B$, where $A\in \np$ and $B \in \sigmathree$. 
Hence $L=L(N^{(A,B)})$.
Define

$C=\{x\condition N^{(A,B)}(x) \mbox{ accepts if }x_A\mbox{ is answered 
correctly and } x_B\mbox{ is answered ``no''}\}$ and 

$D=\{x\condition$
\begin{minipage}[t]{13cm}
$N^{(A,B)}(x)$ after answering the query $x_A$ correctly neither accepts 
nor rejects regardless of the answer to $x_B$, and $x_B \in B\}$.
\end{minipage}

Note that the set $D$ can also be seen as the set of all $x$ such that 
$x_B\in B$ and the partial truth-table of $N^{(A,B)}(x)$ with respect 
to a correct answer to $x_A$ has at least one mind change. 
Clearly, $C \in \p^{\np[1]}$ and $D \in \sigmathree$.
Furthermore, it is not hard to verify that for all $x\in \sigmastar$,
$$x \in L \iff \pair{x,x}\in C\deltatilde D.$$
But note that we also have for all $x \in \sigmastar$, 
$$\pair{x,x}\in C\deltatilde D \iff \pair{f(x),g(x)}\in 
\lnpone\deltatilde\lsigthree,$$ 
where $f$ and $g$ are polynomial-time computable functions 
reducing $C$ to $\lnpone$ and $D$ to $\lsigthree$, respectively.
This shows that $L$ is many-one reducible to $\lnpone\deltatilde\lsigthree$
which completes the proof of the claim.

\item[C] 
Since $\p^{(\p,\sigmathree)} = \p^{(\np,\sigmathree)}$ we have a many-one 
reduction from $\lnpone\deltatilde\lsigthree$ to  
$\lp\deltatilde\lsigthree$. 
In other words, there exists a polynomial-time computable function $h$ 
such that for all $x_1,x_2 \in \sigmastar$,
$$\pair{x_1,x_2} \in \lnpone\deltatilde\lsigthree \iff 
h(\pair{x_1,x_2}) \in \lp\deltatilde\lsigthree.$$
Let $h'$ and $h''$ be the polynomial-time computable functions such that 
for all $x_1,x_2\in \sigmastar$,
$h(\pair{x_1,x_2})=\pair{h'(\pair{x_1,x_2}),h''(\pair{x_1,x_2})}$ and thus 
$$(x_1\in \lnpone \iff x_2 \notin \lsigthree)\iff
(h'(\pair{x_1,x_2})\in \lp \iff h''(\pair{x_1,x_2}) \notin \lsigthree).$$

\item[D] Let $l$ be an integer. 
The string $x_2$ is said to be {\em easy for length $l$\/} 
if and only if 
$(\exists x_1~ |x_1|\leq l)[x_1 \in \lnpone \iff 
h'(\pair{x_1,x_2}) \not\in \lp]$. 

$x_2$ is said to be {\em hard for length $l$\/} if and only if 
$(\forall x_1~ |x_1|\leq l)[x_1 \in \lnpone \iff h'(\pair{x_1,x_2}) \in \lp]$. 

Thus, every string is either easy or hard for length $l$. 
This observation will be used to divide the problem 
of giving a $\sigmathree$ algorithm for $\lsigthreebar$ into two sub-problems, 
which we are going to solve in {\bf E} and {\bf F}. 
Note that testing whether a string $x$ is easy for length $r(|x|)$, 
where $r$ is some polynomial, can be done by a $\sigmatwo$ algorithm.

\item[E] The upcoming Claim~E solves the sub-problem for the strings  
being hard for a certain length.

\item[\phantom{E}]
{\em Claim~E: There exist a set $A \in \sigmathree$ and a polynomial 
$q$ such that for 
all $x \in \sigmastar$, 
if $x$ is hard for length $q(|x|)$ then 
$$ x \not\in \lsigthree \iff x \in A.$$}

Let $p$ be a polynomial such that for all $x \in \sigmastar$,
$$ 
x \not\in \lsigthree \iff (\forall y~|y|\leq p(|x|))(\exists z~|z|\leq p(|x|)
[\pair{x,y,z} \in \lnpone].
$$
Recall that if $x$ is a hard string for length $l$, where $l$ is some 
integer, then
$$(\forall x_1~|x_1|\leq l) [x_1 \in \lnpone \iff  
h'(\pair{x_1,x}) \in \lp].$$ 
Let $q$ be a polynomial such that $q(n)\geq s(p(n))$ for all $n$. 
Suppose that $x$ is a hard string for length $q(|x|)$. 
Hence for all $y,z \in \sigmastar$, $|y|,|z|\leq p(|x|)$, 
$$\pair{x,y,z} \in \lnpone \iff h'(\pair{\pair{x,y,z},x}) \in \lp$$
and thus 
$$ 
x \not\in \lsigthree \iff (\forall y~|y|\leq p(|x|))(\exists z~|z|\leq p(|x|))
[h'(\pair{\pair{x,y,z},x}) \in \lp].
$$
Note that $h'(\pair{v,x})$ is computable in time polynomial in 
$\max\{|v|, |x|\}$. 
Set 
$A=\{x\condition (\forall y~|y|\leq p(|x|))(\exists z~|z|\leq p(|x|))
[h'(\pair{\pair{x,y,z},x}) \in \lp]\}$ and note that $A \in \sigmathree$.

\item[F] We now solve the sub-problem for the strings $x$ being easy 
for length $q(|x|)$.

\item[\phantom{F}] 
{\em Claim~F: Let $q$ be the polynomial defined in {\bf E}. 
There exists a set $B \in \sigmathree$ such that for 
all $x \in \sigmastar$, if $x$ is easy for length $q(|x|)$
then 
$$ x \not\in \lsigthree \iff x \in B.$$}

Define $B=\{x \condition (\exists x_1~ |x_1|\leq q(|x|))
[(x_1 \in \lnpone \iff 
h'(\pair{x_1,x}) \not\in \lp)\wedge h''(\pair{x_1,x}) \in \lsigthree]\}$.
Note that $B \in \sigmathree$.
In light of {\bf C} and {\bf D} this proves the claim.

\item[G]  Combining Claims~E and F with a preliminary test whether 
the input $x$ is hard or easy for length $q(|x|)$, we obtain a $\sigmathree$ 
algorithm for $\lsigthreebar$.

\item[\phantom{G}]
{\em Claim~G: $\lsigthreebar \in \sigmathree$.}

Let $A,B \in \sigmathree$ and $q$ be a polynomial, 
all three as defined in {\bf E} and {\bf F}.
In light of Claims~E and F, the following algorithm is a $\sigmathree$ 
algorithm for $\lsigthreebar$. 
\begin{enumerate}
\item
On input $x$ determine whether the 
input $x$ is easy or hard for length $q(|x|)$. Recall that this can be done 
with one $\sigmatwo$ oracle query according to {\bf D}.
\item
If the input $x$ is hard for length $q(|x|)$ then accept if and only if 
$x \in A$.
\item 
If the input $x$ is easy for length $q(|x|)$ then accept if and only if  
$x \in B$. 

\end{enumerate}

As already pointed out in previous proofs, the use of different oracles in 
the above algorithm does not affect its correctness.

\item[H] Since $\lsigthreebar$ is complete for $\pithree$ we have shown 
$\pithree \subseteq \sigmathree$ and hence $\ph = \sigmathree$. 

\item[\em End of Proof of Main Claim]

\end{description}
\qed

\paragraph{7) Hemaspaandra/Hemaspaandra/Hempel 1996~\cite{hem-hem-hem:tOutByConf:downward-translation,hem-hem-hem:jtoappear:downward-translation}}

\begin{theorem}\label{tt:hhh2}
If $\psigthreetwott=\psigthreethreett$ then 
$\ph = \dsigthree\bolddelta\sigmafour$.
\end{theorem}

\proof
\begin{description}

\item[\em Main Claim]
{\em If $\p_{1,2\hbox{-}{\rm tt}}^{(\p,\sigmathree)} = 
\p_{1,2\hbox{-}{\rm tt}}^{(\np,\sigmathree)}$ then 
$\dsigthree=\co\dsigthree$,}  

where $\p_{1,2\hbox{-}{\rm tt}}^{(\p,\sigmathree)}$ and 
$\p_{1,2\hbox{-}{\rm tt}}^{(\np,\sigmathree)}$ are the classes 
of languages that can be accepted by some DPTM making in parallel at most 
one query to a $\p$ or $\np$ oracle, respectively, and at most two queries 
to a $\sigmathree$ oracle. 
Since $\psigthreetwott \subseteq 
\p_{1,2\hbox{-}{\rm tt}}^{(\p,\sigmathree)} \subseteq 
\p_{1,2\hbox{-}{\rm tt}}^{(\np,\sigmathree)} \subseteq \psigthreethreett$, 
the theorem follows immediately from the above claim 
in light of
Theorem~\ref{t:new}. 
Thus, it suffices to show the correctness of the main claim.

\item[\em Proof of Main Claim:]

\item[A] 
Suppose $\p_{1,2\hbox{-}{\rm tt}}^{(\p,\sigmathree)} = 
\p_{1,2\hbox{-}{\rm tt}}^{(\np,\sigmathree)}$.
Let $\lp$ and $\lnpone$ be many-one complete languages for $\p^{\p[1]}=\p$ and 
$\p^{\np[1]}$, respectively. 
Let $\ldsigthree$ be a $\dsigthree$ complete language. 
We will give a $\dsigthree$ algorithm for $\ldsigthreebar$.
Let $L_1, L_2 \in \sigmathree$ such that $\ldsigthree=L_1-L_2$.

\item[B] {\em  Claim~B: 
$\lp\deltatilde\ldsigthree$ and $\lnpone\deltatilde\ldsigthree$ 
are many-one complete for $\p_{1,2\hbox{-}{\rm tt}}^{(\p,\sigmathree)}$ and 
$\p_{1,2\hbox{-}{\rm tt}}^{(\np,\sigmathree)}$, respectively.}
 
The proof is similar to {\bf B} in the proof of 
Theorem~\ref{tt:hhh1} (Hemaspaandra/Hemaspaandra/Hempel 1996).
But here the $\ldsigthree$ part of $\lnpone\deltatilde\ldsigthree$ 
accounts essentially for 
determining if there is {\em exactly\/} one mind change in the partial 
truth-table with respect to a correctly answered $\np$ query. 

\item[C] Since by assumption $\p_{1,2\hbox{-}{\rm tt}}^{(\p,\sigmathree)} = 
\p_{1,2\hbox{-}{\rm tt}}^{(\np,\sigmathree)}$ it follows that there is 
a many-one reduction $h$ 
between $\lnpone\deltatilde\ldsigthree$ and $\lp\deltatilde\ldsigthree$. 
While continuing as in {\bf C} in the proof of Theorem~\ref{tt:hhh1} 
(Hemaspaandra/Hemaspaandra/Hempel 1996) 
and replacing $\lsigthree$ by $\ldsigthree$ we obtain 
for all $x_1,x_2 \in \sigmastar$,
$$(x_1\in \lnpone \iff x_2 \notin \ldsigthree)\iff
(h'(\pair{x_1,x_2})\in \lp \iff h''(\pair{x_1,x_2}) \notin \ldsigthree).$$

\item[D] As in the proof of Theorem~\ref{tt:hhh1} 
(Hemaspaandra/Hemaspaandra/Hempel 1996).

\item[E] As in the proof of Theorem~\ref{tt:hhh1} 
(Hemaspaandra/Hemaspaandra/Hempel 1996) we are first going to solve the 
sub-problem for the strings being hard for a certain length. 

\item[\phantom{E}]
{\em Claim~E: There exist sets $A_1,A_2 \in \sigmathree$ and 
a polynomial $q$ such that 
for all $x \in \sigmastar$, 
if $x$ is a hard string for length $q(|x|)$ then 
$$x \not\in \ldsigthree \iff x \in A_1-A_2.$$}

A straightforward application of the key idea of  
{\bf E} from the proof of Theorem~\ref{tt:hhh1} 
(Hemaspaandra/Hemaspaandra/Hempel 1996) 
leads to a $\dsigthree$ algorithm to test whether $x \in\ldsigthreebar$ if 
$x$ is a hard string.
Without loss of generality let $p_1$ and $p_2$ be two polynomials such that 
for all $x \in \sigmastar$ and all $i=1,2$,
$$x \in L_i \iff (\exists y~|y|\leq p_i(|x|))(\forall z~|z|\leq p_i(|x|))
[\pair{x,y,z} \in \lnpone].$$ 
Let $q$ be a polynomial such that $q(n)\geq \max\{s(p_1(n)),s(p_2(n))$ 
for all $n$.
Let $x$ be hard for length $q(|x|)$.
Hence we have for all $x_1$, $|x_1|\leq q(|x|)$,
$x_1 \in \lnpone \iff h'(\pair{x_1,x}) \in \lp$.
Hence for all $i=1,2$,
$$
x \in L_i \iff (\exists y~|y|\leq p_i(|x|))(\forall z~|z|\leq p_i(|x|))
[h'(\pair{\pair{x,y,z},x}) \in \lp].
$$
Define for $i=1,2$,
$A'_i=\{x\condition 
(\exists y~|y|\leq p_i(|x|))(\forall z~|z|\leq p_i(|x|))
[h'(\pair{\pair{x,y,z},x}) \in \lp\}$. 
Note that $A'_1,A'_2 \in \sigmatwo$ and that for all $x \in \sigmastar$, 
if $x$ is hard for length $q(|x|)$ then 
$$x \in \ldsigthree \iff x \in A'_1 -A'_2.$$ 
Since $\co{\rm D}\cdot\sigmatwo\subseteq \p^{\sigmatwo[2]} \subseteq 
\dsigthree$ 
there exist sets $A_1, A_2 \in \sigmathree$ such that
$A_1-A_2=\overline{A'_1-A'_2}$.
Hence, 
if $x$ is a hard string for length $q(|x|)$ then 
$x \not\in \ldsigthree \iff x \in A_1-A_2.$

\item[F] Now follows the solution of the sub-problem for the strings $x$ 
being easy for length $q(|x|)$.
\item[\phantom{F}]
{\em Claim~F: 
Let $q$ be the  polynomial defined in {\bf E}.
There exist sets $B_1,B_2 \in \sigmathree$ such that 
for all $x \in \sigmastar$, 
if $x$ is an easy string for length $q(|x|)$ then 
$$x \not\in \ldsigthree \iff x \in B_1-B_2.$$}

Recall $\ldsigthree=L_1-L_2$, where $L_1, L_2 \in \sigmathree$. 
Define for $i=1,2$, 
\begin{center}
$B_i=\{x\condition(\exists x_1~|x_1|\leq q(|x|))$
\begin{minipage}[t]{10cm} 
$[(x_1 \in \lnpone \iff h'(\pair{x_1,x}) \not\in \lp) \wedge$

$\qquad (\forall v <_{lex} x_1)
[v \in \lnpone \iff h'(\pair{v,x}) \in \lp] \wedge$

$\qquad \qquad h''(\pair{x_1,x}) \in L_i]\}$.
\end{minipage}
\end{center}
Obviously, $B_1,B_2 \in \sigmathree$. 
In light of {\bf C} and the definition of $B_1$ and $B_2$, 
it is not hard to verify that 
if $x$ is an easy string for length $q(|x|)$ then
$$
x\not\in \ldsigthree \iff x\in B_1-B_2.
$$

\item[G] Combining the results of {\bf E} and {\bf F} with a 
preliminary test whether the input $x$ is hard or easy for length $q(|x|)$, 
we obtain a $\dsigthree$ algorithm for $\ldsigthreebar$. 

\item[\phantom{G}] 
{\em Claim~G: $\ldsigthreebar \in \dsigthree$.}

Let $A_1,A_2,B_1,B_2 \in \sigmathree$ and $q$ be a polynomial, 
all as defined in {\bf E} and {\bf F}.

For $i=1,2$ let $\wh{L_i}$ be the language accepted by the following algorithm:
\begin{enumerate}
\item 
On input $x$ determine whether the input $x$ is easy or hard 
for length $q(|x|)$. This can be done with one $\sigmatwo$ oracle query 
according to {\bf D}.
\item 
If the input $x$ is hard for length $q(|x|)$ then accept if and only if 
$x \in A_i$.
\item 
If the input $x$ is easy for length $q(|x|)$ accept if and only if 
$x \in B_i$.

\end{enumerate}

Clearly, $\wh{L_1},\wh{L_2}\in \sigmathree$.
Furthermore, for all $x \in \sigmastar$, 
$x \in \ldsigthreebar \iff x \in \wh{L_1}-\wh{L_2}$
due to Claims~E and F.
Hence $\ldsigthreebar \in \dsigthree$.

\item[H] Since $\ldsigthreebar$ is complete for $\co\dsigthree$ we obtain 
$\dsigthree=\co\dsigthree$.

\item[\em End of Proof of Main Claim]

\end{description}
\qed

\paragraph{8) Buhrman/Fortnow 1996~\cite{buh-for:t:two-queries}}

\begin{theorem}\label{tt:bf}
If $\psigtwoone=\psigtwotwo$ then $\ph=\sigmatwo$.
\end{theorem}

\proof
\begin{description}

\item[\em Main Claim] 

{\em If $\psigtwoone=\np\bolddelta\sigmatwo$ then $\ph=\sigmatwo$,}

where $\np\bolddelta\sigmatwo=
\{A\Delta B\condition A\in \np \wedge B\in \sigmatwo\}$.
Since $\psigtwoone \subseteq \np\bolddelta\sigmatwo \subseteq \psigtwotwo$ 
the theorem follows immediately from the above claim.
So we will prove the theorem by proving the main claim.

\item[\em Proof of Main Claim:]

\item[A]
Assume $\psigtwoone=\np\bolddelta\sigmatwo$.
Let $\lp$ and $\lsigtwo$ be complete languages for $\p$ and $\sigmatwo$, 
respectively.
$\lp \deltatilde \lsigtwo$ is complete for $\p^{(\p,\sigmatwo)}=\psigtwoone$.
This can be shown quite analogous to {\bf B} from the proof of 
Theorem~\ref{tt:hhh1} (Hemaspaandra/Hemaspaandra/Hempel 1996). 
Furthermore, observe that $\sat\deltatilde\lsigtwo \in \np\bolddelta\sigmatwo$.

\item[B] Since  
$\psigtwoone = \np\bolddelta\sigmatwo$ 
we have  
$\sat\deltatilde\lsigtwo \in \psigtwoone$.
Consequently there is a many-one reduction $h$ from 
$\sat\deltatilde\lsigtwo$ to $\lp \deltatilde \lsigtwo$.
Continuing as in {\bf C} in the proof of Theorem~\ref{tt:hhh1} 
(Hemaspaandra/Hemaspaandra/Hempel 1996) 
while replacing $\lsigthree$ by $\lsigtwo$ and $\lnpone$ by \sat\ 
yields for all $x_1,x_2 \in \sigmastar$,
$$(x_1\in \sat \iff x_2 \notin \lsigtwo)\iff
(h'(\pair{x_1,x_2})\in \lp \iff h''(\pair{x_1,x_2}) \notin \lsigtwo).$$

\item[C] As in {\bf D} in the proof of Theorem~\ref{tt:hhh1} 
(Hemaspaandra/Hemaspaandra/Hempel 1996) 
but replace $\lnpone$ by \sat.

\item[D] Observe that the proof of Theorem~\ref{tt:hhh1} 
(Hemaspaandra/Hemaspaandra/Hempel 1996) is not valid for $\sigmatwo$ 
instead of $\sigmathree$. 
The crucial point in the proof of Theorem~\ref{tt:hhh1} is the test in the 
final algorithm whether a string is easy or hard for a certain length 
(done by a $\sigmatwo$ oracle query). 
Since the final algorithm of this proof has to be a $\sigmatwo$ algorithm, 
one has to avoid the preliminary easy-hard test  and thus one needs 
to shield each of the sub-algorithms against falsely accepting.

\item[\phantom{D}]
{\em Claim~D: There exist a set $A \in \conp$ and a polynomial 
$q$ such that for all $x \in \sigmastar$, 
\begin{enumerate}
\item 
$A \subseteq \lsigtwobar$ and 
\item 
for all $x\in \sigmastar$, if $x$ is hard for length $q(|x|)$ 
then 
$$x \not\in \lsigtwo \iff x \in A.$$ 
\end{enumerate}}

Without loss of generality let for all $x \in \sigmastar$,
$$x \notin \lsigtwo \iff (\forall z~|z| \leq p(|x|))[\pair{x,z}\in \sat],$$
for some polynomial $p$.
Define $q$ to be a polynomial such that $q(n)\geq s(p(n))$ for all $n$.
Assume that $x$ is a hard string for length $q(|x|)$. 
Hence, 
$$(*) \qquad 
(\forall x_1~|x_1|\leq q(|x|))[x_1 \in \sat \iff h'(\pair{x_1,x}) \in \lp].$$
Consider the following $\conp$ algorithm:
\begin{enumerate}
\item
On input $x$ guess $z$, $|z|\leq p(|x|)$. 
\item 
Assume (*) and use the self 
reduction of $\sat$ to find a potential witness for $\pair{x,z}\in \sat$ with 
the help of (*) in deterministic polynomial time. 
(This is done as follows: Let $\pair{x,z}$ encode the boolean 
formula $F$. We construct $\omega_F$ an assignment for $F$. 
Let the variables in $F$ be ordered. 
Replacing the first variable in $F$ by 0 (1) leads to a boolean formula 
$F_0$ ($F_1$), let the string $v_0$ encode $F_0$. 
Compute $h(\pair{v_0,x})$ and  
test whether $h'(\pair{v_0,x}) \in \lp$. 
If $h'(\pair{v_0,x}) \in \lp$ then under assumption (*) $F_0 \in \sat$. 
Thus, set the value for the first variable in $\omega_F$ to 0 and 
repeat this procedure with $F_0$ until $\omega_F$ assigns a value to each 
variable in $F$. 
If $h'(\pair{v_0,x}) \not\in \lp$ then under assumption (*) 
$F_0 \not\in \sat$ implying 
that a satisfying assignment for $F$, if there exists one, assigns 1 
to the first variable. So set the value for the first variable in $\omega_F$ 
to 1 and repeat this procedure with $F_1$ until $\omega_F$ assigns a value 
to each variable in $F$.)
\item
Accept if and only if the string constructed in step 2 is a witness 
for $\pair{x,z}\in \sat$. (In other words, accept if and only if the 
assignment $\omega_F$ constructed in step 2 satisfies $F$.)
\end{enumerate}

Let $A$ be the language accepted by this algorithm, $A \in \conp$. 
If $x$ is a hard string for length $q(|x|)$ then 
(*) in fact holds and it is not hard to verify that
$$x\not\in \lsigtwo \iff x \in A.$$
But note that even if $x$ is not a hard string for length $q(|x|)$,
$x \in A \implies  x\not\in \lsigtwo.$
This follows from the fact that the algorithm only accepts 
(as it is a $\conp$ algorithm) 
if for all 
$z$, $|z|\leq p(|x|)$, the string constructed in step 2 is a 
witness for $\pair{x,z} \in \sat$.

\item[E] The sub-algorithm for the strings $x$ being easy for length $q(|x|)$ 
given in {\bf F} from the proof of 
Theorem~\ref{tt:hhh1} (Hemaspaandra/Hemaspaandra/Hempel 1996)
has already the required shielding feature as spoken of in the beginning 
of {\bf D} and can be easily adapted to the current proof. 

\item[\phantom{E}]
{\em Claim~E: 
Let $q$ be the polynomial defined in {\bf D}.
There exists a set $B\in \sigmatwo$ such that 
\begin{enumerate}
\item $B \subseteq \lsigtwobar$ and 
\item for all 
$x\in \sigmastar$, if $x$ is an easy string for length $q(|x|)$ then
$$x \not\in \lsigtwo \iff x\in B.$$
\end{enumerate}
}  

It is not hard to see that Claim~E can be shown quite analogous to 
{\bf F} from the proof of 
Theorem~\ref{tt:hhh1} (Hemaspaandra/Hemaspaandra/Hempel 1996)  
with respect to the obvious adaptions as for instance a replacement 
of $\lsigthree$ by $\lsigtwo$ and $\lnpone$ by $\sat$.  
Hence, for the modified set $B$ from {\bf F} in the proof of 
Theorem~\ref{tt:hhh1} (Hemaspaandra/Hemaspaandra/Hempel 1996) 
holds, 
$x \in B$ if and only if 
$(\exists x_1~|x_1|\leq q(|x|))[(x_1 \in \sat \iff 
h'(\pair{x_1,x})\not\in \lp) \wedge h''(\pair{x_1,x}) \in \lsigtwo]$.
Note that $x \in B$ if and only if 
$x$ is easy for length $q(|x|)$ and $x\not\in \lsigtwo$.
In particular, if $x$ is hard for length $q(|x|)$ then $x \not\in B$.

\item[F] Running the sub-algorithms from Claims~D and E in parallel gives 
a $\sigmatwo$ algorithm for $\lsigtwobar$. 

\item[\phantom{F}]
{\em Claim~F: $\lsigtwobar \in \sigmatwo$.}

Let $A\in \conp$, $B\in \sigmatwo$, and $q$ be a polynomial, all three 
as defined in {\bf D} and {\bf E}. 
We have the following $\sigmatwo$ algorithm for $\lsigtwobar$.
\begin{enumerate}
\item 
On input $x$ guess whether the string $x$ is hard or easy  
for length $q(|x|)$.
\item
If the algorithm has guessed that $x$ is a hard string for length $q(|x|)$ 
accept if and only if $x \in A$.
\item 
If ``$x$ is an easy string for length $q(|x|)$'' was guessed in step 1 then 
accept if and only if $x \in B$. 
\end{enumerate}

Recall from {\bf D} and {\bf E} that the sub-algorithm emerging from the wrong 
guess does not accept if $x \not\in \lsigtwobar$.

\item[G] Since $\lsigtwobar$ is many-one complete for $\pitwo$ we obtain 
$\pitwo=\sigmatwo$ and thus $\ph = \sigmatwo$.

\item[\em End of Proof of Main Claim]

\end{description}
\qed

\paragraph{9) Hemaspaandra/Hemaspaandra/Hempel 1997~\cite{hem-hem-hem:t:translating-downwards}}

\begin{theorem}
If $\psigtwotwott=\psigtwothreett$ then 
$\ph = \dsigtwo\bolddelta\sigmathree$.
\end{theorem}

\proof
\begin{description}

\item[\em Main Claim] 
{\em If $\p\bolddelta\dsigtwo=\np\bolddelta\dsigtwo$ then 
$\dsigtwo=\co\dsigtwo$,}

where $\p\bolddelta\dsigtwo$ is defined similar to $\np\bolddelta\dsigtwo$ 
as in {\bf Main Claim} of the proof of 
Theorem~\ref{tt:bf} (Buhrman/Fortnow 1996).
Since $\psigtwotwott \subseteq \p\bolddelta\dsigtwo
\subseteq \np\bolddelta\dsigtwo \subseteq \psigtwothreett$ 
the theorem follows immediately from the above claim in light 
of Theorem~\ref{t:new}. 
It remains to prove the main claim.

\item[\em Proof of Main Claim:]

 \item[A]
Suppose $\p\bolddelta\dsigtwo=\np\bolddelta\dsigtwo$.
Let $\lp$, $\lnp$, and $\ldsigtwo$ be complete languages for $\p$, $\np$, 
and  $\dsigtwo$, respectively. 
Let $L_1,L_2 \in \sigmatwo$ such that $\ldsigtwo=L_1-L_2$.

\item[B] {\em Claim~B:} 
{\em $\lp\deltatilde\ldsigtwo$ and $\lnp\deltatilde\ldsigtwo$  
are many-one complete for $\p\bolddelta\dsigtwo$ and $\np\bolddelta\dsigtwo$, 
respectively.}

The proof is straightforward and thus omitted.

\item[C] Since $\p\bolddelta\dsigtwo=\np\bolddelta\dsigtwo$, 
$\lnp\deltatilde\ldsigtwo$ many-one reduces to 
$\lp\deltatilde\ldsigtwo$. 
Continue as in {\bf C} in the proof of Theorem~\ref{tt:hhh1} 
(Hemaspaandra/Hemaspaandra/Hempel 1996) 
but replace $\lsigthree$ by $\ldsigtwo$ and $\lnpone$ by $\lnp$.
Thus for all $x_1,x_2 \in \sigmastar$,
$$(x_1\in \lnp \iff x_2 \notin \ldsigtwo)\iff
(h'(\pair{x_1,x_2})\in \lp \iff h''(\pair{x_1,x_2}) \notin \ldsigtwo).$$

\item[D] As {\bf D} in the proof of Theorem~\ref{tt:hhh1} 
(Hemaspaandra/Hemaspaandra/Hempel 1996) 
but replace $\lnpone$ by $\lnp$.

\item[E] As {\bf D} of the proof of Theorem~\ref{tt:bf} (Buhrman/Fortnow 1996) 
but replace $\lsigtwo$ by $L_1$.

\item[F] The result of {\bf E} can be extended to yield a $\sigmatwo$ 
algorithm for the hard strings $x$ for length $q(|x|)$ in $\ldsigtwobar$ 
that is protected against accepting if the input string $x$ in fact is 
an easy string for length  $q(|x|)$ and $x \not\in \lsigtwobar$.

\item[\phantom{F}]{\em Claim~F: Let $q$ be the polynomial defined in {\bf E}. 
There exists a set $A' \in \sigmatwo$ such that 
\begin{enumerate}
\item 
$A' \subseteq \ldsigtwobar$ and 
\item 
for all $x \in \sigmastar$, 
if $x$ is a hard string for length $q(|x|)$ then 
$$x \not\in \ldsigtwo \iff x \in A'.$$ 
\end{enumerate}
}

Let $L_1,L_2 \in \sigmatwo$ and $A \in \conp$ be as defined in 
{\bf A} and {\bf E}. 
Define $A'= A \cup L_2$.
Clearly, $A'  \in \sigmatwo$. 
Note that  
\begin{eqnarray*}
x \in A'  & \iff &  x \in A \vee x\in L_2 \\
& \Longrightarrow & x \in \overline{L_1} \vee x\in L_2 \\  
& \iff & x \in \overline{L_1} \cup L_2 \\
& \iff & x \notin L_1-L_2\\
& \iff & x \in \ldsigtwobar.
\end{eqnarray*}
Hence $A' \subseteq \ldsigtwobar$. 
Furthermore, in case $x$ is hard for length 
$q(|x|)$ the second line in the above implication chain turns  
into an equivalence according to {\bf E}.

\item[G] There is also a sub-algorithm for the strings $x$ 
being easy for length $q(|x|)$.
Observe that a straightforward adaption of {\bf F} from the proof of 
Theorem~\ref{tt:hhh1} (Hemaspaandra/Hemaspaandra/Hempel 1996) 
does not work here, since the sets $B_i$ constructed there would 
(even with the adaption to the current situation) remain 
$\sigmathree$ sets, but one needs $\sigmatwo$ sets.

\item[\phantom{G}]
{\em Claim~G: 
Let $q$ be the polynomial defined in {\bf E}.
There exist sets $B_1,B_2 \in \sigmatwo$ such that,
\begin{enumerate}
\item 
$B_1-B_2 \subseteq \ldsigtwobar$ and 
\item 
for all $x \in \sigmastar$, if $x$ is an easy string for length $q(|x|)$ then 
$$x \not\in \ldsigtwo \iff x \in B_1-B_2.$$
\end{enumerate}
} 

Recall that $\ldsigtwo=L_1-L_2$ where $L_1, L_2 \in \sigmatwo$. 
Without loss of generality let 
$L_1 \superseteq L_2$~\cite{cai-gun-har-hem-sew-wag-wec:j:bh1}.
Define for $i=1,2$,
$$B_i=\{x\condition (\exists x_1~|x_1|\leq q(|x|))
[(x_1 \in \lnp \iff h'(\pair{x_1,x}) \not\in \lp) 
\wedge h''(\pair{x_1,x}) \in L_i]\}.$$
Note that $B_1, B_2 \in \sigmatwo$ and $B_1 \superseteq B_2$. 
We will prove the claim by showing that for all $x \in \sigmastar$, 
$x \in B_1-B_2$ if and only if $x$ is easy for length $q(|x|)$ and 
$x \not\in L_1-L_2$.

Let $x \in \sigmastar$.
Observe that for all $i=1,2$,
$x \in B_1-B_2$ implies $x$ is easy for length $q(|x|)$.
So it suffices to show that if $x$ is easy for length $q(|x|)$ then 
$x \not\in \ldsigtwo \iff x \in B_1-B_2$.

So let $x$ be easy for length $q(|x|)$.
Let $t=\max(\{0\}\cup\{i\in \{1,2\}\condition x\in B_i\})$. 
Let $z_2$  be a string such that 
\begin{center}
$(\exists x_1~|x_1|\leq q(|x|))$
\begin{minipage}[t]{7cm}  
$[z_2=h''(\pair{x_1,x}) \wedge$

$\qquad (x_1 \in \lnp \iff  h'(\pair{x_1,x}) \not\in \lp) \wedge$

$\qquad \qquad (t>0 \implies z_2 \in L_t)]$.
\end{minipage}
\end{center}
Such a string $z_2$ exists since $x$ is easy for length $q(|x|)$. 
Note that $x \not\in L_1-L_2 \iff z_2 \in L_1-L_2$. 
This follows from the definition of $z_2$ and the fact the equivalence
$$(x_1\in \lnp \iff x \notin \ldsigtwo)\iff
( h'(\pair{x_1,x}) \in \lp \iff  h''(\pair{x_1,x}) \notin \ldsigtwo)$$ 
does hold for all $x_1 \in \sigmastar$ according to {\bf C}.
Furthermore, $x \in B_1-B_2$ if and only if  $z_2 \in L_1-L_2$ due to the 
definition of $z_2$, $B_1$, $B_2$, and $t$. 
Thus, 
\begin{eqnarray*}
x\in\ldsigtwobar & \iff & x \not\in L_1-L_2\\
& \iff & z_2 \in  L_1-L_2\\
& \iff & x \in B_1-B_2.\\
\end{eqnarray*}

\item[H] Combining Claims~F and G while exploiting the structure of 
$\dsigtwo$ shows

\item[\phantom{H}]{\em Claim~H: $\ldsigtwobar \in \dsigtwo$.}

Let the sets $A',B_1,B_2 \in \sigmatwo$ be as defined in {\bf F} and {\bf G}.
We show the above claim by proving 
$\ldsigtwobar=(B_1 \cup A' )-B_2$.

Suppose $x \in \ldsigtwobar$. 
Note that $x$ is either easy or hard for length $q(|x|)$.
If $x$ is easy for that length then $x \in B_1-B_2$ according to Claim~G.
If $x$ is hard for length $q(|x|)$ then $x \in A'$ 
according to Claim~F  and $x\notin B_2$ according to Claim~G.
In both cases we certainly have $x \in (B_1 \cup A' )-B_2$.

Now suppose $x \in (B_1 \cup A' )-B_2$. 
Hence $x \in B_1 \cup A'$. 
If $x \in A'$ then $x\in \ldsigtwobar$ according to Claim~F.
If  $x \not\in A'$ then $x\in B_1-B_2$. 
But this implies $x \in \ldsigtwobar$ according to Claim~G. 

\item[H] Since $\ldsigtwobar$ is complete for $\co\dsigtwo$ we obtain 
$\dsigtwo=\co\dsigtwo$.

\item[\em End of Proof of Main Claim]

\end{description}
\qed

\section{Open Questions and Literature Starting Points}
\label{s:questions}

In our final section we would like to address some very interesting open 
questions related to the topic of this survey. 
Can one show 
$$\psigkjqueries=\psigkjplusone \implies \ph=\psigkjqueries$$
for $k\geq 1$ and $j\geq 2$? 
Note that the above claim for $k>2$ and $j=1$ has been proven 
in~\cite{hem-hem-hem:tOutBydown-sep:upward-sep} whereas the claim 
for $k=2$ and $j=1$ first appeared in~\cite{buh-for:t:two-queries}.

Can one prove a downward collapse result within the bounded-truth-table 
hierarchy and the boolean hierarchy over $\np$?
In particular, for which, if any, $j$ does it hold that 
$$\p^{\np}_{j\hbox{-}{\rm tt}}=\p^{\np}_{(j+1)\hbox{-}{\rm tt}} 
\implies \diff_j(\np)={\rm co}\diff_j(\np)?$$
Buhrman and Fortnow~\cite{buh-for:t:two-queries} give an oracle relative to 
which this fails for $j=1$. 
But for $j=2,3,\dots$ not even that is known. 
Note that the $\np$ case seems special; for $\sigmak$, $k\geq 2$, 
the relation holds for all $j$ (by~\cite{hem-hem-hem:jtoappear:downward-translation,buh-for:t:two-queries,hem-hem-hem:t:translating-downwards}).

Hemaspaandra and Rothe~\cite{hem-rot:j:boolean} study boolean 
hierarchies over UP, 
and Bertoni et al.~\cite{ber-bru-jos-sit-you:c:gen} study boolean 
hierarchies over R.
Do the techniques developed in the papers~\cite{hem-hem-hem:jtoappear:downward-translation,buh-for:t:two-queries,hem-hem-hem:t:translating-downwards} have 
any application to the unambiguous polynomial hierarchy, or to the 
``R hierarchy''?

Finally, if one wants to go into the actual journal literature, where 
should one start?  
Perhaps the best place in the original
literature to get a first feel for the easy-hard technique 
in its initial form is 
Kadin's paper~\cite{kad:joutdatedbychangkadin:bh}, which 
initiated the technique.
Perhaps the best place in the original
literature to get a feel for the easy-hard technique in
the more powerful ``downward collapse'' form 
is the paper 
of Hemaspaandra, Hemaspaandra,
and Hempel~\cite{hem-hem-hem:jtoappear:downward-translation}, which
initiated this form of the technique.  Finally, for completeness,
we mention that strongest currently known downward translations
regarding the research line 
described in this survey 
are found in the final 
paper surveyed,~\cite{hem-hem-hem:t:translating-downwards}, except
that very recently 
the key result of one subpart of that paper has itself 
been further extended in~\cite{hem-hem-hem:t:diff-diff-downwards}.

\bibliography{gry}

\end{document}